\documentclass[%
	10pt,%
	amsmath,%
	amsfonts,%
	amssymb,%
	aps,
	prl,%
	twocolumn,%
	superscriptaddress,
	longbibliography
]{revtex4-1}
\usepackage[utf8]{inputenc}
\usepackage[T1]{fontenc}
\usepackage[english]{babel}
\usepackage{csquotes}

\usepackage[%
  colorlinks=true,%
	citecolor=blue,%
	linkcolor=black,%
	urlcolor=blue%
]{hyperref}

\usepackage{graphicx}
\graphicspath{{Figures/}}

\usepackage{bm}
\usepackage{upgreek}

\def\ifnonemptyparenthesis#1{%
  \if\relax\detokenize{#1}\relax%
  \else%
    (#1)%
  \fi%
}

\newcommand{\Ztwo}{$\mathbb{Z}_{2}$}
\newcommand{\ii}{\mathrm{i}}
\newcommand{\e}{\mathrm{e}}
\newcommand{\s}{\mathrm{s}}
\newcommand{\hc}{\mathrm{h.c.}}
\newcommand{\ket}[1]{\left|#1\right>}
\newcommand{\bra}[1]{\left<#1\right|}
\newcommand{\ketone}{\left|1\right>}
\newcommand{\braone}{\left<1\right|}
\newcommand{\kettwo}{\left|2\right>}
\newcommand{\bratwo}{\left<2\right|}
\newcommand{\ketthree}{\left|3\right>}
\newcommand{\brathree}{\left<3\right|}
\newcommand{\ketfour}{\left|4\right>}
\newcommand{\brafour}{\left<4\right|}
\newcommand{\bessel}[2][]{\mathcal{J}_{#1}(#2)}
\newcommand{\besselchi}[1]{\mathcal{J}_{#1}(\chi)}
\newcommand{\fig}[2][]{Fig.\,\ref{#2}\ifnonemptyparenthesis{#1}}

\newcommand{\invisiblesection}[1]{%
  \phantomsection%
  \stepcounter{section}%
  \addcontentsline{toc}{section}{\protect\numberline{\thesection}#1}%
}

\begin{document}
%%%%%%%%%%%%%%%%% Title %%%%%%%%%%%%%%%%%%%%%%%%%%%%%%%%%%%%%%%%%%%%%%%
\title{Floquet approach to $\mathbb{Z}_{2}$ lattice gauge theories
with ultracold atoms in optical lattices}

\author{Christian Schweizer}
\affiliation{Fakult\"at f\"ur Physik, 
             Ludwig-Maximilians-Universit\"at, 
             Schellingstr.\ 4, D-80799 M\"unchen, Germany}
\affiliation{Max-Planck-Institut f\"ur Quantenoptik, 
             Hans-Kopfermann-Strasse 1, D-85748 Garching, Germany}
\affiliation{Munich Center for Quantum Science and Technology (MCQST),
             Schellingstr. 4, D-80799 M\"unchen, Germany}
\author{Fabian Grusdt}      
\affiliation{Department of Physics, Technical University of Munich, 
             D-85748 Garching, Germany}
\affiliation{Munich Center for Quantum Science and Technology (MCQST), 
             Schellingstr. 4, D-80799 M\"unchen, Germany}
\author{Moritz Berngruber}
\affiliation{Fakult\"at f\"ur Physik, 
             Ludwig-Maximilians-Universit\"at, 
             Schellingstr.\ 4, D-80799 M\"unchen, Germany}
\affiliation{Munich Center for Quantum Science and Technology (MCQST),
             Schellingstr. 4, D-80799 M\"unchen, Germany}
\author{Luca Barbiero}
\affiliation{Center for Nonlinear Phenomena and Complex Systems, 
             Universit\'e Libre de Bruxelles, CP 231, 
						 Campus Plaine, B-1050 Brussels, Belgium}
\author{Eugene Demler}
\affiliation{Department of Physics, Harvard University, Cambridge, 
             Massachusetts 02138, USA}
\author{Nathan Goldman}
\affiliation{Center for Nonlinear Phenomena and Complex Systems, 
             Universit\'e Libre de Bruxelles, CP 231, 
						 Campus Plaine, B-1050 Brussels, Belgium}
\author{Immanuel Bloch}
\affiliation{Fakult\"at f\"ur Physik, 
             Ludwig-Maximilians-Universit\"at, 
             Schellingstr.\ 4, D-80799 M\"unchen, Germany}
\affiliation{Max-Planck-Institut f\"ur Quantenoptik, 
             Hans-Kopfermann-Strasse 1, D-85748 Garching, Germany}
\affiliation{Munich Center for Quantum Science and Technology (MCQST),
             Schellingstr. 4, D-80799 M\"unchen, Germany}
\author{Monika Aidelsburger}
\affiliation{Fakult\"at f\"ur Physik, 
             Ludwig-Maximilians-Universit\"at, 
             Schellingstr.\ 4, D-80799 M\"unchen, Germany}
\affiliation{Max-Planck-Institut f\"ur Quantenoptik, 
             Hans-Kopfermann-Strasse 1, D-85748 Garching, Germany}
\affiliation{Munich Center for Quantum Science and Technology (MCQST),
             Schellingstr. 4, D-80799 M\"unchen, Germany}

\maketitle

%%%%%%%%%%%%%%%%% Absctract %%%%%%%%%%%%%%%%%%%%%%%%%%%%%%%%%%%%%%%%%%%
\textbf{
Quantum simulation
has the potential to investigate gauge theories 
in strongly-interacting regimes,
which are up to now inaccessible through conventional numerical techniques.
Here, we take a first step in this direction by implementing a Floquet-based method
for studying $\mathbb{Z}_2$ lattice gauge theories
using two-component ultracold atoms in a double-well potential.
For resonant periodic driving at the on-site interaction strength
and an appropriate choice of the modulation parameters, the 
effective Floquet Hamiltonian exhibits $\mathbb{Z}_2$ symmetry. 
We study the dynamics of the system for different initial states 
and critically contrast the observed evolution with a theoretical analysis 
of the full time-dependent Hamiltonian of the periodically-driven lattice model. 
We reveal challenges that arise due to symmetry-breaking terms 
and outline potential pathways to overcome these limitations. 
Our results provide important insights for future studies of lattice gauge theories 
based on Floquet techniques.
}

%%%%%%%%%%%%%%%%% Introduction %%%%%%%%%%%%%%%%%%%%%%%%%%%%%%%%%%%%%%%%
\invisiblesection{Introduction}
Lattice gauge theories (LGTs)~\cite{Wilson:1974ji,Kogut:1979wg} 
are fundamental for our understanding 
of quantum many-body physics across different disciplines ranging from
condensed matter~\cite{Wen2004,Levin:2005js,Lee:2006de,Ichinose:2014ko} 
to high-energy physics~\cite{Aoki:2017}.  
However, theoretical studies of LGTs can be extremely challenging 
in particular in strongly-interacting regimes,
where conventional computational methods are 
limited~\cite{Troyer:2005hv,Alford:2008ft}.
To overcome these limitations alternative numerical tools
are currently developed, which enable out-of-equilibrium 
and finite density computations~\cite{Buyens:2016jf,Banuls:2016cc,PietroSilvi:2017ja,Gazit:2017cn}. 
In parallel, the rapid progress in the field of 
quantum simulation~\cite{Weimer:2010ez,Blatt:2012gw,Gross:2017do,Romero:2016vm}
has sparked a growing interest in designing experimental platforms
to explore the rich physics of 
LGTs~\cite{Tagliacozzo:2013,Wiese:2013kka,Zohar:2015gs,Dalmonte:2016hd,Notarnicola:2015ee,Kasper:2017gs,Kuno:2017,Zhang:2018}. 
State-of-the-art experiments are now able to explore the physics 
of static~\cite{Aidelsburger:2018cr} 
as well as density-dependent gauge fields~\cite{Clark:2018fi} and have 
engineered controlled few-body interactions~\cite{Anderlini:2007jg,Trotzky:2008jy,Dai:2017en}, 
which are the basis for many proposed schemes to realize LGTs.
First studies of the Schwinger model have been performed with
quantum-classical algorithms~\cite{Klco:2018} 
and a digital quantum computer composed of 
four trapped ions~\cite{Martinez:2016gx}.
The challenge for analog quantum simulators mainly lies
in the complexity to engineer gauge-invariant interactions 
between matter and gauge fields.

Here, we explore the dynamics of a minimal model
for $\mathbb{Z}_2$~LGTs coupled to matter
with ultracold atoms in periodically-driven double-well potentials~\cite{Barbiero:2018wg}. 
An alternative technique was recently proposed for digital 
quantum simulation~\cite{Zohar:2017}.
$\mathbb{Z}_2$~LGTs are of high interest 
in condensed matter physics~\cite{Horn:1979,Ju:2013er,Gazit:2017cn,GonzalezCuadra:2018vk} 
and topological quantum computation~\cite{Kitaev:2003ul}.
Our scheme is based on density-dependent laser-assisted tunneling techniques~\cite{Keilmann:2011,Greschner:2015,Bermudez:2015,Straeter:2016}.
We use a mixture of bosonic atoms in two different internal states 
to encode the matter and gauge field degrees of freedom. 
The interaction between these states is engineered via resonant 
periodic modulation~\cite{Goldman:2015kca,Goldman:2014,Bukov:2015gu,Eckardt:2017hca} 
of the on-site potential at the inter-species Hubbard interaction~\cite{Ma:2011fr,Chen:2011dn,Meinert:2016ky,Gorg:2018de}. By choosing suitable modulation parameters, the 
effective Floquet model exhibits a $\mathbb{Z}_2$ symmetry~\cite{Barbiero:2018wg}.
We present a detailed study of this effective Floquet model defined on a double well,
which constitutes the basic building block of the LGT. 
We discuss the relation between the observed dynamics and the ideal model
and reveal the potential impact of symmetry-breaking terms.

\begin{figure}[t!]
\includegraphics{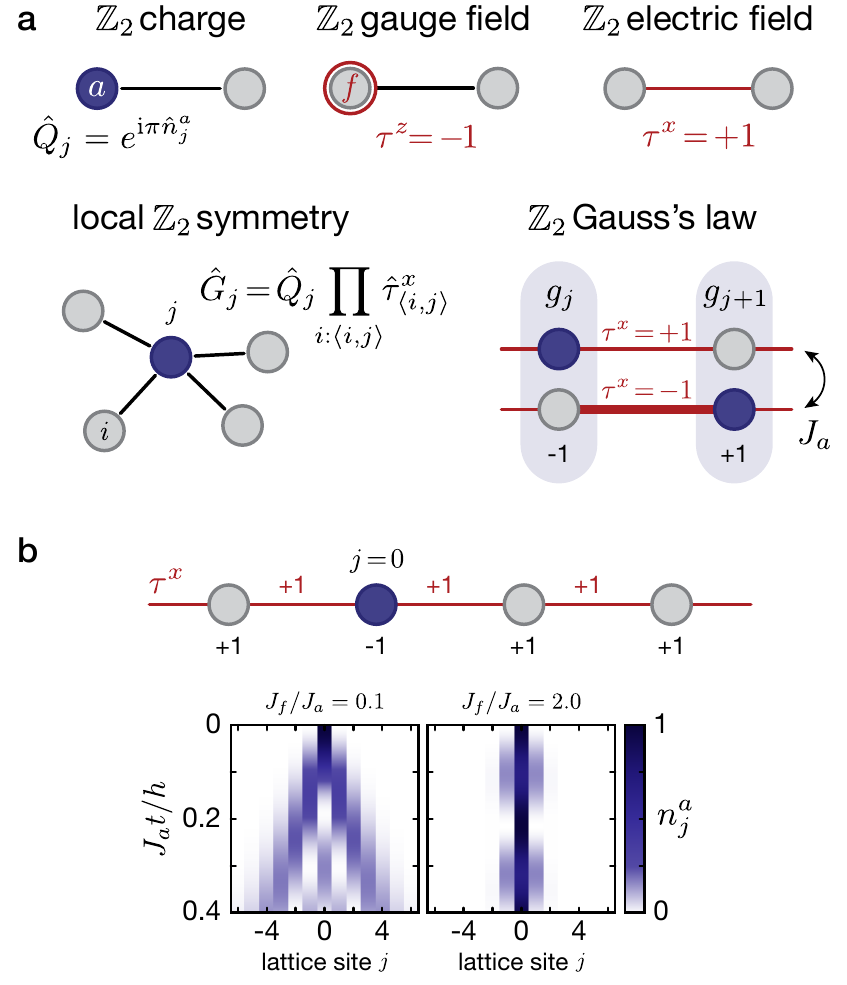}
\caption{\textbf{1D $\mathbb{Z}_2$ lattice gauge theory coupled to matter.}
Circles indicate lattice sites, which are empty (gray) 
or occupied by a matter particle (blue). 
Red circles and the thickness of red links illustrate the
expectation value of the link operators, $\tau^z$ and $\tau^x$.
\textbf{a} Elementary ingredients: $\mathbb{Z}_2$ charge $\hat{Q}_j\!=\!\text{e}^{i\pi\hat{n}_j^a}$, 
$\mathbb{Z}_2$ gauge field $\hat{\tau}^z_{\langle j, j+1 \rangle }$, 
$\mathbb{Z}_2$ electric field $\hat{\tau}^x_{\langle j, j+1 \rangle }$, 
and local symmetry operator $\hat{G}_j$ with conserved quantities~$g_j$. Here $i\!:\!\langle i,j \rangle$ denotes all lattice sites $i$ connected to site $j$ via a nearest-neighbor link, denoted as $\langle i,j \rangle$.
Matter and gauge fields are implemented using two different species, denoted as $a$ (blue) and $f$ (red). 
Matter-gauge coupling occurs with strength $J_a$.
\textbf{b} Dynamics of the 1D model~\eqref{eq:H1D} for different values of $J_f/J_a$
calculated with exact diagonalization 
of a system with 13 sites based on Eq.~\eqref{eq:H1D}.
The initial state is a single matter particle 
located on site $j\!=\!0$ and the gauge field is in an eigenstate of the electric field.
}%
\label{fig:1}%
\end{figure}

In order to understand the observed phenomena, it is instructive to consider the 
properties of an extended one-dimensional (1D) $\mathbb{Z}_2$ LGT,
as captured by the Hamiltonian
\begin{align}
\hat{H}_{\mathbb{Z}_2} \!=\! &- \sum\limits_j J_a \left( \hat{\tau}_{\langle j, j+1 \rangle }^z \hat{a}_{j}^\dagger \hat{a}_{j+1}^{\phantom\dagger} 
   +\text{h.c.}\right) \nonumber\\
   &- \sum\limits_j J_f \hat{\tau}_{\langle j, j+1 \rangle }^x.
    \label{eq:H1D}
\end{align}
Here $\hat{a}_{j}^{\dagger}$ describes the creation 
of a matter particle on lattice site~$j$ 
and the Pauli operators ${\bf\hat{\tau}}_{\langle j, j+1 \rangle }$, 
defined on the links between neighboring lattice sites, encode the gauge field degrees of freedom. 
The elementary ingredients of this $\mathbb{Z}_2$~LGT are illustrated in Fig.~\ref{fig:1}a. 
Note that the illustrations of the $\mathbb{Z}_2$ gauge field and $\mathbb{Z}_2$ electric field 
are related to the physical implementation of the building block, 
which is discussed later in the text.
The matter field has a charge $\hat{Q}_j\!=\!\text{e}^{i\pi\hat{n}_j^a}$ on site $j$, 
which is given by the parity of the site occupation, 
with $\hat{n}_j^a\!=\!\hat{a}_{j}^{\dagger}\hat{a}_j^{\phantom\dagger}$ the number operator.
The dynamics of the matter field is coupled to 
the $\mathbb{Z}_2$~gauge field $\hat{\tau}^z_{\langle j, j+1 \rangle }$ with an amplitude $J_a$.
The energy scale associated with the electric field $\hat{\tau}^x_{\langle j, j+1 \rangle }$ is $J_f$.

The model Hamiltonian~\eqref{eq:H1D} commutes with the lattice gauge transformations 
defined by the local symmetry operators 
\begin{align}
\hat{G}_j \!=\! \hat{Q}_j \prod_{i:\langle i, j \rangle } \hat{\tau}_{\langle i, j \rangle}^x, 
            \qquad \left[ \hat{H}, \hat{G}_j \right]\!=\!0\ \  \forall j,
\label{eq:gop}
\end{align}
\noindent where $\prod_{i:\langle i, j \rangle }$ 
denotes the product over all nearest-neighbor links connected to lattice site~$j$. 
The eigenvalues of $\hat{G}_j$ are $g_j\!=\!\pm 1$. 
The dynamics of the model is constrained by $\mathbb{Z}_2$ Gauss's law,
$\hat{G}_j |\psi \rangle\!=\!g_j |\psi\rangle$, 
in analogy to electrodynamics.
Since the local values $g_j$ are conserved, the motion of $\mathbb{Z}_2$ charges is
coupled to a change of the $\mathbb{Z}_2$ electric field lines on the link connecting the two lattice sites.
Gauss's law effectively separates the Hilbert space into different subsectors, 
which are characterized by a set of conserved quantities $\{ g_j\}$.
The two configurations sketched in Fig.~\ref{fig:1}a (lower right)
belong to the same subsector and illustrate the basic matter-gauge 
coupling according to Gauss's law.
Lattice sites with $g_j\!=\!-1$ are interpreted as local static background charges (Supplementary Information).
Different subsectors can be explored by preparing suitable initial states.

In order to gain more insight into the physics of the 1D model~\eqref{eq:H1D},
we consider a system initially prepared in an eigenstate 
of the electric field operator,
with $\tau^x\!=\!+1$ on all links, 
and a single matter particle located on site~$j=0$. 
For this initial state $g_j\!=\!+1$, $\forall j\neq 0$ and 
$g_{j}\!=\!-1$ for $j\!=\!0$ (Fig.~\ref{fig:1}b). 
In the limit of vanishing electric field energy $J_f\rightarrow0$, 
the matter particle can tunnel freely 
along the 1D chain, thereby changing the electric field on all traversed links. 
For $J_f \neq 0$, tunneling of the matter particle is detuned 
due to the energy of the electric field and the matter particle is 
bound to the location of the static background charge at $j=0$.
In this regime the energy of the system scales linearly 
with the distance between the static charge 
and the matter particle, which we interpret as a signature
of confinement (Supplementary Information).

%%%%%%%%%%%%%%%%% Implementation %%%%%%%%%%%%%%%%%%%%%%%%%%%%%%%%%%%%%%
\invisiblesection{Implementation of the two-site model}
\begin{figure*}[ht]%
\includegraphics[width=\textwidth]{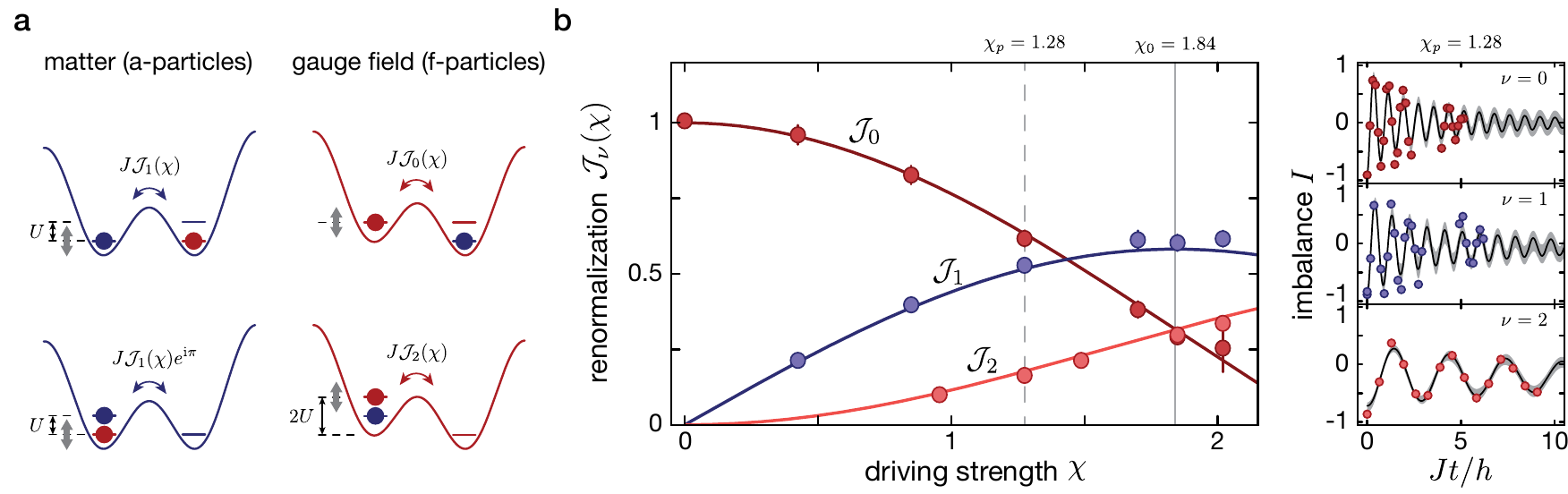}%
\caption{\textbf{Driving scheme for $\mathbb{Z}_2$ LGTs on a double well.}
\textbf{a} Effective tunneling processes 
for the matter field (blue, $a$) and gauge field (red, $f$) particle. 
For $\phi=0$, hopping of $a$-particles occurs for resonant one-photon processes at $\hbar\omega\!\approx\!U$ 
with an effective amplitude $J\mathcal{J}_{1}(\chi)$, where
$U$ is the inter-species on-site interaction. 
Depending on the position of the $f$-particle, the $a$-particle acquires a phase shift of $\pi$, 
which realizes the matter-gauge coupling. 
Tunneling of the $f$-particle is renormalized by zero- or induced via two-photon processes,
with amplitudes $J\mathcal{J}_0(\chi)$ and $J\mathcal{J}_2(\chi)$ depending on the $a$-particle's position.
\textbf{b} Experimental results for the renormalization of the tunnel couplings $\mathcal{J}_{\nu}(\chi)$
for single-particle $\nu$-photon processes $\nu\!=\!\{0,1,2\}$, 
with $\omega \!=\! 2\pi \times 4122\,$Hz and $J/h\!\approx\! 0.5\,$kHz.
The solid lines are the Bessel functions, 
where $\chi$ was calibrated by fitting the zeroth-order Bessel function 
to the dark red data points (Supplementary Information).
The time traces of the imbalance $I$
are fitted with sinusoidal functions taking into account
an inhomogeneous tilt distribution (solid black line)
and shown for exemplary traces at $\chi_p\!=\!1.28$ 
(dashed vertical line, left panel) on the right. 
The error bars and the gray shading are the $1\sigma$-confidence interval
obtained from a bootstrap analysis of 1000 repetitions (Supplementary Information).
The solid gray vertical line marks the value $\chi_0$, 
where $\mathcal{J}_0(\chi_0)\!=\!\mathcal{J}_2(\chi_0)$.}%
\label{fig:2}%
\end{figure*}

Here, we engineer the elementary interactions of the $\mathbb{Z}_2$ model 
on a two-site lattice following Ref.~\cite{Barbiero:2018wg}. 
The matter and gauge fields are implemented using two different species 
denoted as $a$- and $f$-particles,
which are realized by two Zeeman levels of the hyperfine ground-state manifold of $^{87}$Rb, 
$\left|a\right>\equiv\left|F\!=\!1,m_F\!=\!-1\right>$ and
$\left|f\right>\equiv\left|F\!=\!1,m_F\!=\!+1\right>$. 
We prepare one $a$- and one $f$-particle
in each two-site system.
The matter field is associated with the $a$-particle. 
The $\mathbb{Z}_2$ gauge field is the number imbalance 
$\hat{\tau}^z_{\langle j, j+1 \rangle }\!=\!\hat{n}^f_{j+1}-\hat{n}^f_{j}$ of the $f$-particle
and the $\mathbb{Z}_2$ electric field corresponds to tunneling of the $f$-particle, 
$\hat{\tau}^x_{\langle j, j+1 \rangle }\!=\!\hat{f}_{j}^{\dagger}\hat{f}_{j+1}^{\phantom\dagger}+\hat{f}_{j+1}^{\dagger}\hat{f}_{j}^{\phantom\dagger}$, where $\hat{f}_{j}^{\dagger}$ is the creation 
operator of an $f$-particle on site~$j$ 
and $\hat{n}^f_j\!=\!\hat{f}_{j}^{\dagger}\hat{f}_{j}^{\phantom\dagger}$ 
is the corresponding number-occupation operator. 
An extension of our scheme to realize extended 1D LTGs is presented in the Supplementary Information.
It requires exactly one $f$-particle per link, 
while the density of $a$-particles
(fermions or hard-core bosons) can take arbitrary values.

The driving scheme is based on a species-dependent double-well potential 
with tunnel coupling~$J$ between neighboring sites 
and an energy offset $\Delta_f$ only seen by the $f$-particle.
Experimentally, it is realized with a magnetic-field gradient, 
making use of the opposite magnetic moments of the two states 
$|a\rangle$ and $|f\rangle$ (Supplementary Information).
In the limit of strong on-site interactions $U\gg J$, first-order tunneling processes are suppressed
but can be restored resonantly with a periodic modulation 
at the resonance frequency $\hbar\omega=\sqrt{U^2+4J^2}\approx U$.
The full time-dependent Hamiltonian can be expressed as 
\begin{equation}
\begin{aligned}
\hat{H}(t) \!=\!
   &-J \left(\hat{a}^\dagger_2 \hat{a}^{\phantom\dagger}_1 
     + \hat{f}^\dagger_2 \hat{f}^{\phantom\dagger}_1 
     + \textrm{h.c.}\right) \\
   &+ U \sum\limits_{j\!=\!1,2}\hat{n}^{a}_j \hat{n}^{f}_j + \Delta_f \hat{n}_1^f \\
   &+ A \cos{\left(\omega t + \phi\right)}\,(\hat{n}^a_1 + \hat{n}^f_1)
   ,
\end{aligned}
\label{eq:Ht}
\end{equation}
where $A$ is the modulation amplitude and $\phi$ is the modulation phase. 
For resonant modulation $\hbar \omega\approx U$ and in the high-frequency limit $\hbar \omega \gg J$, 
the lowest order of the
effective Floquet Hamiltonian contains renormalized tunneling matrix elements
for both $a$- and $f$-particles~\cite{Goldman:2014,Bukov:2015gu,Eckardt:2017hca}.
For general modulation parameters,
the amplitudes and phases are operator-valued and explicitly depend on the 
site-occupations.
For certain values of the modulation phase ($\phi\!=\!0$ or $\pi$), however,  
these expressions simplify and realize the $\mathbb{Z}_2$ model. 
The driving scheme can be understood by considering the 
individual photon-assisted tunneling processes of $a$- and $f$-particles in
situations, where one of the two particles is localized on a particular site of the double well. 
This generates an occupation-dependent energy offset for the other particle, which is equal 
to the on-site Hubbard interaction $U$ (Fig.~\ref{fig:2}a).
For all configurations tunneling is resonantly restored for energy differences 
$\nu \hbar \omega$ between neighboring sites
with renormalized tunneling $J \mathcal{J}_{\nu}(\chi)\text{e}^{i\nu\phi}$; here
$\nu$ is an integer,
$\mathcal{J}_{\nu}$ is the
$\nu$th-order Bessel function of the first kind and 
$\chi\!=\!A/(\hbar\omega)$ the dimensionless driving parameter.

For $\phi=0$, we find that the strength of $a$-particle tunneling
is density-independent $J_a = J |\mathcal{J}_{\pm 1}(\chi)|$,
however, depending on the position of the $f$-particle, 
the on-site energy difference between neighboring sites is either $+ U$ or $- U$ (Fig.~\ref{fig:2}a).
This results in a sign-dependence of the renormalized tunneling $\pm J_a$, 
which stems from the property 
of odd Bessel functions
$\mathcal{J}_{-\nu}(\chi)\!=\!(-1)^{\nu}\mathcal{J}_{\nu}(\chi)$ (Supplementary Information)
and is central to our implementation 
of the $\mathbb{Z}_2$ symmetry~\cite{Barbiero:2018wg}. 
It allows us to write the renormalized hopping of 
$a$-particles as $J_{a}\,\hat{\tau}^z_{\langle 1,2 \rangle}$. 
Note that we drop the link-index from now on to simplify notations,
${\bf \hat{\tau}}\equiv{\bf\hat{\tau}}_{\langle 1,2 \rangle}$.

Tunneling of $f$-particles becomes real-valued,
with an amplitude that only weakly depends on the position of the $a$-particle.
Due to the species-dependent tilt~$\Delta_f=U$ the on-site energy difference between neighboring sites
is either $\Delta_f\!-\!U\!=\!0$ or $\Delta_f\!+\!U\!=\!2U$ (Fig.~\ref{fig:2}a).
Therefore, tunneling is renormalized via zero- and two-photon processes,
resulting in the real-valued tunneling matrix elements $J\mathcal{J}_0(\chi)$ 
and $J\mathcal{J}_2(\chi)$. 
To lowest order, the effective double-well Hamiltonian takes the form
\begin{equation}
\hat{H}_{\text{eff}} \!=\!
    -J_{a}\,\hat{\tau}^z
    \left(\hat{a}^\dagger_2 \hat{a}^{\phantom\dagger}_1
    + \hat{a}^\dagger_1 \hat{a}^{\phantom\dagger}_2\right)
    - \hat{J}_{f}\,\hat{\tau}^x
    ,
    \label{eq:DWeff}
\end{equation}
where $\hat{J}_{f}$ depends on the position of the $a$-particle
\begin{equation}
\hat{J}_{f} \!=\! 
    J \mathcal{J}_0(\chi)\,\hat{n}^{a}_1 
    + J \mathcal{J}_2(\chi)\,\hat{n}^{a}_2 .
\end{equation}
The density dependence of $\hat{J}_{f}$ can be avoided 
by choosing the dimensionless driving strength~$\chi$ such
that $\mathcal{J}_0(\chi_{0})\!=\!\mathcal{J}_2(\chi_{0})$,
which occurs, e.g.\ at $\chi_{0}\!\approx\!1.84$. 
Then, Eq.~\eqref{eq:DWeff} reduces to the two-site version
of the $\mathbb{Z}_2$ LGT described by Hamiltonian~\eqref{eq:H1D}.
Note that the double-well model defined in Eq.~\eqref{eq:DWeff} 
is $\mathbb{Z}_2$-symmetric for all values of the driving strength $\chi$.

\begin{figure}%
\includegraphics{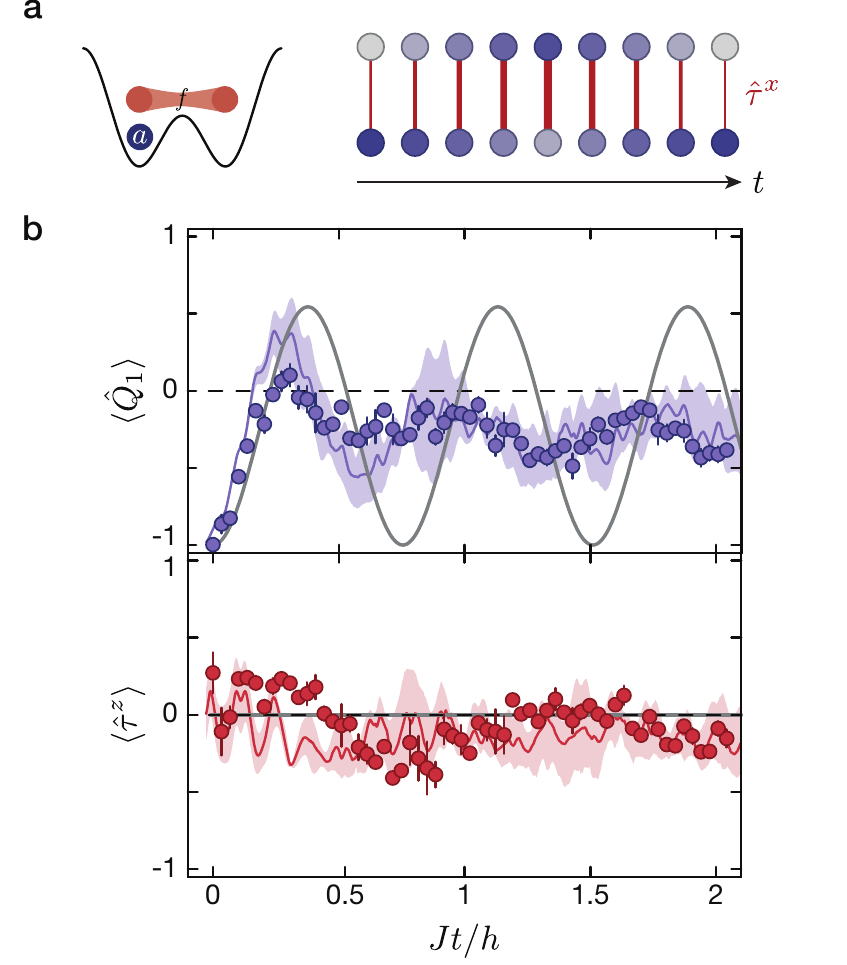}%
\caption{\textbf{Dynamics of the matter-gauge system prepared initially in an 
eigenstate of the electric field~$\hat{\tau}^x$.}
\textbf{a} Illustration of the gauge invariant initial state 
$\left| \psi_0^x\right>\!=\!|a, 0\rangle \otimes \left(|f, 0 \rangle + | 0,f\rangle\right)/\sqrt{2}$
and the expected dynamics according to Hamiltonian~\eqref{eq:DWeff}.
\textbf{b} Measured expectation values of the $\mathbb{Z}_{2}$~charge $\langle \hat{Q}_1 \rangle$ (blue points) 
and $\mathbb{Z}_{2}$~gauge field $\langle\hat{\tau}^z\rangle$ (red points) for $\omega\!=\!2\pi\!\times\!4320\,$Hz.
Each data point represents the mean of at least three individual experimental results and
the error bars denote the standard deviation. 
The blue and red lines and shadings show a numerical analysis using time-dependent 
exact-diagonalization, which includes averaging of the 
observables in the presence of an inhomogeneous tilt distribution $\Delta(x,y,z)$
approximated by a normal distribution with standard deviation
$\Delta_\sigma/h \!=\! 0.44(2)\,$kHz, which was independently calibrated (Supplementary Information).
The blue and red solid line is the median and the shading represents 
the $1\sigma$-confidence interval obtained with a bootstrap analysis of 1000 repetitions.
All calculations are performed using the independently calibrated experimental parameters, 
$J/h \!=\! 587(3)\,$Hz, $\Delta_f/h\!=\!4.19(3)\,$kHz, $U/h \!=\! 3.85(7)\,$kHz and 
taking into account additional terms that appear in the extended Bose-Hubbard model (Supplementary Information). 
The gray solid lines are the ideal dynamics
according to Eq.~\eqref{eq:DWeff} and Eq.~\eqref{eq:ideal}.
}%
\label{fig:3}%
\end{figure}

The experimental setup consists of a 3D optical lattice 
generated at wavelength $\lambda_s\!=\!767\,$nm. 
Along the $x$-axis an additional standing wave with wavelength $2\lambda_s\!=\!1534\,$nm 
is superimposed to create a superlattice potential. 
For deep transverse lattices and suitable superlattice parameters, 
an array of isolated double-well potentials is realized, 
where all dynamics is restricted to the two double-well sites (Supplementary Information).
The periodic drive is generated by modulating the amplitude 
of an additional lattice with wavelength $2\lambda_s$, 
whose potential maxima are aligned relative to the double-well potential 
in order to modulate only one of the two sites. 
This enables the control of the modulation phase, 
which is set to $\phi\!=\!0$ or $\pi$.

We first study the renormalization of the tunneling matrix elements 
for the relevant $\nu$-photon processes~\cite{Keay:1995jo,Lignier:2007,Sias:2008jf,Mukherjee:2015cr,Meinert:2016ky}
with a single atom on each double well (\fig{fig:2}b).
For every measurement, 
the atom is initially localized on the lower-energy site
with a potential energy difference $\Delta_\nu\!\approx\! \nu \hbar \omega$
to the higher-energy site, where $\nu \in \{0, 1, 2\}$. 
Then, the resonant modulation is switched on rapidly at frequency~$\omega$ 
and we evaluate the imbalance $I\!=\!n_2-n_1$ as a function of the evolution time, 
where $n_j$ is the density on site $j$.
These densities were determined using 
site-resolved detection methods~\cite{Trotzky:2008jy}.
Note, this technique provides an average of this observable over the entire 3D array 
of double-well potentials.
Hence, an overall harmonic confinement 
and imperfect alignment of the lattice laser beams introduces
an inhomogeneous tilt distribution $\Delta (x,y,z)$,
which leads to dephasing of the averaged dynamics.
The renormalized tunneling amplitude is obtained from the oscillation frequency 
of the imbalance and by numerically taking into account the tilt distribution $\Delta (x,y,z)$~[Fig.~\ref{fig:2}b]. 
We find that our data agrees well with the expected Bessel-type behavior 
for the $\nu$-photon processes (Supplementary Information). 
Moreover, these measurements enable us to directly determine 
the value of the modulation amplitude,
for which $\mathcal{J}_0(\chi_{0})\!=\!\mathcal{J}_2(\chi_{0})$,
as indicated by the vertical line in Fig.~\ref{fig:2}b.

\begin{figure}%
\includegraphics{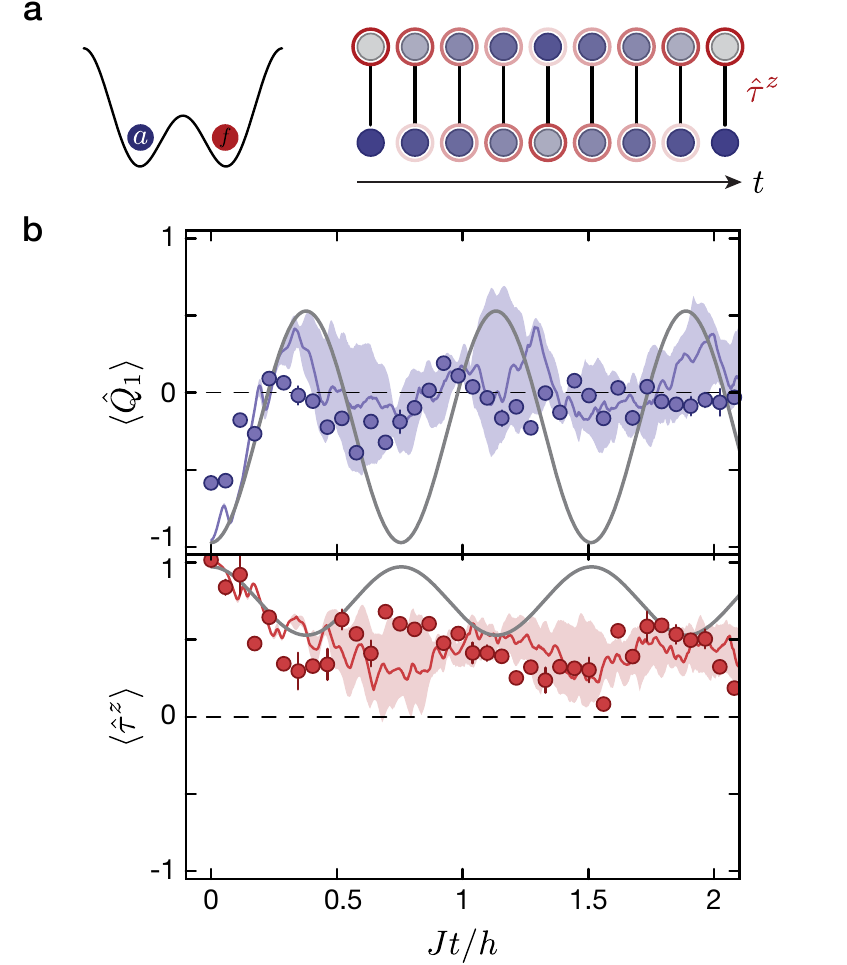}%
\caption{\textbf{Dynamics of the matter-gauge system prepared initially in an eigenstate of the gauge field $\hat{\tau}^z$.}
\textbf{a} Schematic of the initial state 
$\left| \psi_0^z\right>\!=\!|a, 0\rangle \otimes |0, f \rangle$
and the expected dynamics.
Brightness of the red circles illustrate the expectation value of $\hat{\tau}^z$.
\textbf{b} Measured expectation values of the $\mathbb{Z}_{2}$~charge $\langle \hat{Q}_1 \rangle$ (blue points) 
and the $\mathbb{Z}_{2}$~gauge field $\langle\hat{\tau}^z\rangle$ (red points) for $\omega\!=\! 2\pi\!\times\!4314\,$Hz.
Each data point represents the mean of at least three individual experimental results and
the error bars denote the standard deviation. 
The blue and red lines and shadings show a numerical analysis using time-dependent 
exact-diagonalization, with $J/h \!=\! 578(3)\,$Hz, $\Delta_f/h \!=\! 4.19(3)\,$kHz, 
$U/h \!=\! 3.85(7)\,$kHz and $\Delta_\sigma/h \!=\! 0.46(2)\,$kHz 
as explained in the caption of Fig.~(\ref{fig:3}b) and the Supplementary Information.
The blue and red solid line is the median and the shading represents 
the $1\sigma$-confidence interval obtained with a bootstrap analysis of 1000 repetitions.
The gray solid lines are the ideal dynamics
according to Eq.~\eqref{eq:DWeff} and Eq.~\eqref{eq:ideal}.
}%
\label{fig:4}%
\end{figure}

%%%%%%%%%%%%%%%%% Experimental results %%%%%%%%%%%%%%%%%%%%%%%%%%%%%%%%
\invisiblesection{Experimental results}
In order to study the dynamics of the $\mathbb{Z}_2$ double-well model \eqref{eq:DWeff},
we prepare two different kinds of initial states, where the gauge field particle is either
prepared in an eigenstate of the electric field $\hat{\tau}^x$ (Fig.~\ref{fig:3}) 
or the gauge field operator $\hat{\tau}^z$ (Fig.~\ref{fig:4}a).
In both cases the matter particle 
is initially localized on site $j\!=\!1$.

First, we consider the state 
$\left| \psi_0^x\right> \!=\! |a, 0\rangle \otimes \left(|f, 0 \rangle + | 0,f\rangle\right)/\sqrt{2}$ (Fig.~\ref{fig:3}a),
where the gauge field particle is in a symmetric superposition between the two sites.
This state is an eigenstate of $\hat{G}_j$ defined in Eq.~\eqref{eq:gop}. 
The corresponding eigenvalues are $g_1\!=\!-1$ and $g_2\!=\!+1$. 
After initiating the dynamics by suddenly turning on the resonant modulation, 
we expect that the matter particle starts to tunnel to
the neighboring site ($j\!=\!2$) according to the matter-gauge coupling.
Depending on the energy of the electric field $J_f$, 
this process can be energetically detuned and the matter particle does not 
fully tunnel to the other site. Solving the dynamics 
according to Hamiltonian~\eqref{eq:DWeff} analytically, gives:
\begin{align}
\langle \hat{Q}_1 (t)\rangle &\!=\! - \frac{J_f^2+J_a^2 \cos\left(2 t \sqrt{J_f^2+J_a^2}\right)}{J_f^2+J_a^2}. 
\label{eq:ideal}
\end{align}
The maximum value of $\langle \hat{Q}_1 \rangle $ is limited
to $(J_a^2 - J_f^2)/(J_a^2 + J_f^2)$. 
The experimental configuration is well suited to explore the regime
$J_f/J_a=\mathcal{J}_0(\chi_0)/\mathcal{J}_1(\chi_0)\!\approx\!0.54$, 
which corresponds to an intermediate regime
between the two limiting cases discussed in Fig.~\ref{fig:1}c.
These cases can also be understood at the level of the two-site model.
In the weak electric field regime ($J_f/J_a \ll 1$) the matter particle
tunnels freely between the two sites, while in the limit of a strong electric field
($J_f/J_a \gg 1$) the matter particle remains
localized.

In the experiment we can directly access 
the value of the charge operator $\hat{Q}_j\!=\!\text{e}^{i\pi\hat{n}_j^a}$ 
and the link operator $\hat{\tau}^z\!=\!\hat{n}^f_{2}-\hat{n}^f_{1}$ via 
site- and state-resolved detection techniques~\cite{Trotzky:2008jy}.
They provide direct access to the state-resolved density 
on each site of the double well $n_j^a$ 
and $n_j^f$, averaged over the entire 3D array of double-well realizations.
The experimental results are shown in Fig.~\ref{fig:3}b for $U/J \!=\! 6.6$
and $\phi\!=\!0$. As expected, we find that the charge oscillates,
while the dynamics of the $f$-particle is strongly suppressed.
We observe a larger characteristic oscillation frequency for the $a$-particle compared
to the prediction of Eq.~\eqref{eq:ideal} [gray line, Fig.~\ref{fig:3}b]. This is
predominantly caused by an inhomogeneous tilt distribution $\Delta(x,y,z)$ present in our system.
Taking the inhomogeneity into account, the numerical analysis of the full time-dependence 
according to Eq.~\eqref{eq:Ht} [solid blue line, Fig.~\ref{fig:3}b]
shows good agreement with the experimental results.
The fast oscillations both in the data and the numerics 
are due to the micromotion at non-stroboscopic times.

The $f$-particle is initially prepared in an eigenstate 
of the electric field operator $\hat{\tau}^x$, 
which corresponds to an equal superposition of the particle on both sites 
of the double-well potential, i.e.\ $\langle\hat{\tau}^z (t\!=\!0)\rangle\!=\!0$.
The $\mathbb{Z}_2$ electric field follows the
oscillation of the matter particle in a correlated manner to 
conserve the local quantities $g_j$.
At the same time the expectation value of the gauge field $\langle\hat{\tau}^z (t)\rangle$ 
is expected to remain zero at all times. 
This is a non-trivial result, which is a direct consequence of the $\mathbb{Z}_2$-symmetry
constraints. In contrast, a resonantly driven double-well system with $\Delta_f\!=\!0$,
which does not exhibit $\mathbb{Z}_2$ symmetry, would show 
dynamics with equal oscillation amplitudes for the $a$- and $f$-particle. 
In the experiment we clearly observe suppressed dynamics for the $f$-particle, 
which is a signature of the experimental realization of the 
$\mathbb{Z}_2$ symmetry (Fig.~\ref{fig:3}b).
Deviations between the time-dependent numerical analysis and the experimental results are
most likely due to an imperfect initial state, residual energy offsets, and finite ramp times.

In a second set of experiments we study the dynamics where the gauge field particle is initialized in
an eigenstate of the gauge field operator $\hat{\tau}^z$, while the matter particle
is again localized on site $j\!=\!1$, $\left| \psi_0^z\right> \!=\! |a, 0\rangle \otimes |0, f \rangle$ (Fig.~\ref{fig:4}a).
Here, the system is in a coherent superposition of the two subsectors with $g_1=-g_2\!=\!\pm1$
and the expectation value of the locally conserved operators are 
$\langle \hat{G}_1\rangle\!=\!\langle \hat{G}_2\rangle\!=\!0$. 
Note that there is no coupling between different subsectors according to Hamiltonian~\eqref{eq:DWeff}.
The basic dynamics can be understood in the two limiting cases of the model.
For $J_f\ll J_a$ the electric field vanishes and the system is dominated by the gauge field $\hat{\tau}^z$. 
In this limit, a system prepared in an eigenstate of $\hat{\tau}^z$ will remain in this eigenstate
because $\hat{\tau}^z$ commutes with Hamiltonian~\eqref{eq:DWeff} for $J_f\!=\!0$.
In the opposite regime ($J_f\!\gg\!J_a$), where the electric field dominates, $\langle\hat{\tau}^z\rangle$ oscillates
between the two eigenvalues.
The dynamics of the $\mathbb{Z}_2$ charge on the other hand is still determined by Eq.~\eqref{eq:ideal}.
In the experiment we probe the intermediate regime 
at $J_f/J_a\!\approx\!0.54$ for $U/J \!=\! 6.7$ and $\phi\!=\!\pi$ (Fig.~\ref{fig:4}b). 
The dynamics agrees with the ideal evolution (gray line, Fig.~\ref{fig:4}b) for short times. 
For longer times it deviates due to the averaging over the
inhomogeneous tilt distribution $\Delta(x,y,z)$, which is well captured by the full time-dynamics
according to Hamiltonian~\eqref{eq:Ht} [red and blue lines, Fig.~\ref{fig:4}b].
Notably, $\langle\hat{\tau}^z\rangle$ exhibits a non-zero average value.

%%%%%%%%%%%%%%%%% Discussion %%%%%%%%%%%%%%%%%%%%%%%%%%%%%%%%%%%%%%%%%%
\invisiblesection{Discussion of corrections to the ideal case}

\begin{figure}[t]%
\includegraphics{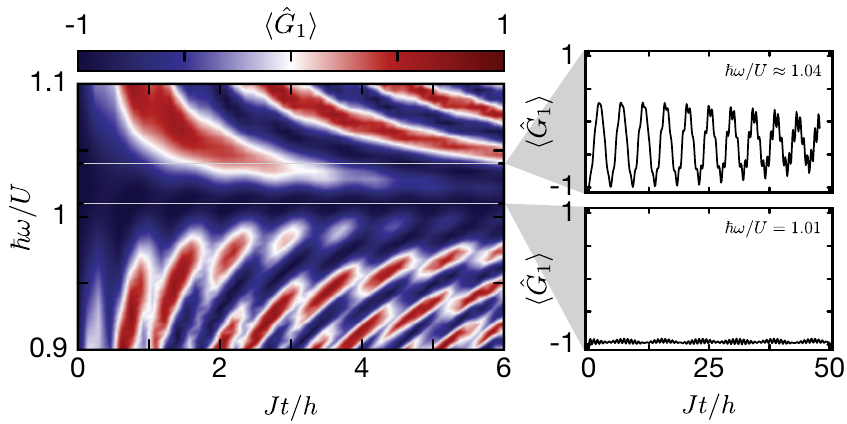}%
\caption{\textbf{Finite-frequency corrections to the effective Floquet Hamiltonian~\eqref{eq:DWeff}.}
Stroboscopic dynamics of the expectation value 
of the local $\mathbb{Z}_{2}$-symmetry operator $\langle \hat{G}_1 \rangle$
for $|\psi^x_0\rangle$, with $g_1=-1$, based
on Hamiltonian~\eqref{eq:Ht} for different driving frequencies $\omega$.
The panels on the right show examples of the time traces 
for $\hbar\omega \!\approx\! 1.04\,U$ and $\hbar\omega \!=\! 1.01\,U$.
}%
\label{fig:5}%
\end{figure}

An important requirement for quantum simulations of gauge theories is the exact implementation
of the local symmetry constraints in order to assure that $\langle \hat{G}_j\rangle$ is conserved for all times.
Since we do not have direct access to $\langle \hat{G}_j\rangle$ experimentally, we study the
implications of symmetry-breaking terms numerically.
The dominant contribution stems from the inhomogeneous tilt distribution $\Delta(x,y,z)$, 
which, however, can be avoided in future experiments
by generating homogeneous box potentials.
The second type of gauge-variant terms are coupling processes that do not fulfill the constraints of Gauss's law. 
These are correlated two-particle tunneling
and nearest-neighbor interactions, which are known to exist in interacting lattice models~\cite{Scarola:2005}. 
For our experimental parameters these terms
are on the order of $0.03\,J$ 
and can be neglected for the timescale of the observed dynamics (Supplementary Information).
The same gauge-variant processes appear in the higher-order terms 
of the Floquet expansion for finite $U/J$
~\cite{Goldman:2014,Bukov:2015gu,Eckardt:2017hca}.
We study them analytically
by calculating the first-order terms of the expansion and 
comparing them to numerics (Supplementary Information).
In Fig.~\ref{fig:5} we show the numerically-calculated dynamics for the initial state 
$|\psi^x_0\rangle$ (Fig.~\ref{fig:3}a) according to the
full time-dependent Hamiltonian [Eq.~\eqref{eq:Ht}] for $U/J\!=\!7$,
similar to the experimental values (Fig.~\ref{fig:5}).
We find that in this regime the driving frequency is crucial
and defines the timescale 
for which the $\mathbb{Z}_2$ symmetry of the model remains.
In particular there is an optimal value around $\hbar\omega\!\approx\!1.01\,U$, 
where the value of $\langle\hat{G}_1\rangle$ deviates by $<10\%$ 
even for long evolution times.

%%%%%%%%%%%%%%%%% Summary %%%%%%%%%%%%%%%%%%%%%%%%%%%%%%%%%%%%%%%%%%%%%
\invisiblesection{Summary}
In summary, we have studied the dynamics of a minimal model for $\mathbb{Z}_2$ LGTs.
Our observations are well described by a full time-dependent analysis of the 3D system.
Moreover, we find non-trivial dynamics of the matter and gauge field
in agreement with predictions from the ideal $\mathbb{Z}_2$ LGT.
We further reached a good understanding of relevant symmetry-breaking terms.
The dominant processes we identified are species-independent energy offsets between neighboring sites
and correlated two-particle tunneling terms~\cite{Scarola:2005},
which can be suppressed in future experiments.
We have further provided important insights into the applicability of Floquet schemes. 
While the Floquet parameters can be fine-tuned in certain cases to ensure gauge invariance,
this complication can be avoided by reaching the high-frequency limit% 
~\cite{Goldman:2014,Bukov:2015gu,Eckardt:2017hca} to minimize finite-frequency corrections. 
In experiments this could be achieved using Feshbach resonances 
to increase the inter-species scattering length.
This, however, comes at the cost of enhanced correlated tunneling processes,
which in turn can be suppressed by increasing the lattice depth (Supplementary Information).
Numerical studies further indicate 
that certain experimental observables are robust to gauge-variant imperfections~\cite{Banerjee:2013,Kuhn:2014},
which may facilitate future experimental implementations.
We anticipate that the double-well model demonstrated in this work
serves as a stepping stone for experimental studies
of $\mathbb{Z}_2$ LGTs coupled to matter
in extended 1D and 2D systems,
which can be realized by coupling many double-well links
along a 1D chain (Supplementary Information) or in a ladder configuration~\cite{Barbiero:2018wg}.
Finally, the use of state-dependent optical lattices
could further enable an independent tunability 
of the matter- and gauge-particle tunneling terms. 

\subsection{Acknowledgements}
We acknowledge insightful discussions with M.~Dalmonte, A.~Trombettoni and M.~Lohse. 
This work was supported by the Deutsche Forschungsgemeinschaft 
(DFG, German Research Foundation) 
under project number 277974659 via Research Unit FOR 2414 
and under project number 282603579 via DIP,
the European Commission (UQUAM Grant No.\ 5319278) and 
the Nanosystems Initiative Munich (NIM, Grant No.\ EXC4). 
The work was further funded by the Deutsche Forschungsgemeinschaft (DFG, German
Research Foundation) under Germany's Excellence Strategy -- EXC-2111 -- 390814868.
Work in Brussels was supported by the FRS-FNRS (Belgium) 
and the ERC Starting Grant TopoCold.
F.~G.\ additionally acknowledges support by the Gordon and Betty Moore foundation 
under the EPIQS program and from the Technical University of Munich -- 
Institute for Advanced Study, funded by the German Excellence Initiative 
and the European Union FP7 under grant agreement 291763, 
from the DFG grant No.\ KN 1254/1-1, and DFG TRR80 (Project F8).
F.~G.\ and E.~D.\ acknowledge funding from Harvard-MIT CUA, 
AFOSR-MURI Quantum Phases of Matter (grant FA9550-14-1-0035),
AFOSR-MURI:\ Photonic Quantum Matter (award FA95501610323) and
DARPA DRINQS program (award D18AC00014).

\subsection{Author contributions}
C.S., F.G., N.G. and M.A. planned the experiment and performed theoretical calculations. 
C.S. and M.B. performed the experiment and analyzed the data with M.A. 
All authors discussed the results and contributed to the writing of the paper.

\subsection{Data availability}
The data that support the plots within this paper and other findings of this study are available from the corresponding author upon reasonable request.

\subsection{Code availability}
The code that supports the plots within this paper are available from the corresponding author upon reasonable request.

%%%%%%%%%%%%%%%%% Bibliography %%%%%%%%%%%%%%%%%%%%%%%%%%%%%%%%%%%%%%%%

\onecolumngrid\newpage

\renewcommand{\thefigure}{S\the\numexpr\arabic{figure}-10\relax}
 \setcounter{figure}{10}
\renewcommand{\theequation}{S.\the\numexpr\arabic{equation}-10\relax}
 \setcounter{equation}{10}
 \renewcommand{\thesection}{S.\Roman{section}}
\setcounter{section}{10}
\renewcommand{\bibnumfmt}[1]{[S#1]}
\renewcommand{\citenumfont}[1]{S#1}

\onecolumngrid
\clearpage
\begin{center}
\noindent\textbf{Supplementary Information for:}
\\\bigskip
%%%%%%%%%%%%%%%%% Title %%%%%%%%%%%%%%%%%%%%%%%%%%%%%%%%%%%%%%%%%%%%%%%
\noindent\textbf{\large{%
       Floquet approach to $\mathbb{Z}_2$ lattice gauge theories with ultracold atoms in optical lattices}}
\\\bigskip
Christian Schweizer,$^{1,2,3}$ Fabian Grusdt,$^{4,3}$
Moritz Berngruber,$^{1,3}$ Luca Barbiero,$^{5}$ Eugene Demler,$^{6}$\\
Nathan Goldman,$^{5}$ Immanuel Bloch,$^{1,2,3}$ and Monika Aidelsburger$^{1,2,3}$\\\vspace{0.3em}
\small{$^{1}$\,\emph{Fakult\"at f\"ur Physik, 
             Ludwig-Maximilians-Universit\"at, 
             Schellingstr.\ 4, D-80799 M\"unchen, Germany}}\\
\small{$^{2}$\,\emph{Max-Planck-Institut f\"ur Quantenoptik, 
             Hans-Kopfermann-Strasse 1, D-85748 Garching, Germany}}\\
\small{$^{3}$\,\emph{Munich Center for Quantum Science and Technology (MCQST), 
             Schellingstr.\ 4, 
             D-80799 M\"unchen, Germany}}\\
\small{$^{4}$\,\emph{Department of Physics, Technical University of Munich, 
             D-85748 Garching, Germany}}\\
\small{$^{5}$\,\emph{Center for Nonlinear Phenomena and Complex Systems,\\
             Universit\'e Libre de Bruxelles, CP 231,
             Campus Plaine, B-1050 Brussels, Belgium}}\\
\small{$^{6}$\,\emph{Department of Physics, Harvard University, Cambridge, 
             Massachusetts 02138, USA}}
\end{center}
\bigskip
\bigskip
\twocolumngrid

\section{Interpretation: static local background charges}
Below Eq.~\eqref{eq:gop} in the main text 
we introduced the notion of ``static local background charges'' for lattice sites, where $g_j=-1$.
Here, we want to briefly explain this in a more formal manner. 
In this interpretation, we consider a second, 
immobile species $b$ of matter particles (corresponding to another particle flavor), 
which are localized on the sites with $g_j=-1$. 
If a single $b$ particle is localized on site~$j$, 
it contributes to the $\mathbb{Z}_2$ charge 
$\hat{\tilde{Q}}_j = \exp[\ii\pi(\hat{n}^a_j + \hat{n}^b_j)]$,
where $\hat{n}^b_j$ is the number operator for $b$ particles. 

In particular, for the dynamics presented in Fig.~1b,
this interpretation results in the following scenario: 
We initialize a single $a$-particle on the central site $j=0$ 
and the field in an eigenstate of the electric field, $\tau^x=1$, on all links. 
This results in $g_j=1$ on all sites except at $j=0$, where $g_0=-1$. 
Consequently, there is a static background charge localized at $j=0$. 
The $\mathbb{Z}_2$ Gauss's law with the charge operator $\hat{\tilde{Q}}_j$ given above, 
then results in $\tilde{g}_j = 1$ for all~$j$,
which is conventionally the physical subsector.
Moreover, when the $a$-particle moves, 
it is connected by a string of $\mathbb{Z}_2$ electric-field lines to the localized $b$ particle. 
The energy cost of this electric field string leads to a linear string tension, 
which can be interpreted as a simple instant of confinement.

\section{Effective Hamiltonian}
\subsection{Floquet expansion}
The \Ztwo{} double-well model realized in this work
consists of a two-site potential 
with one $a$- and one $f$-particle 
implementing the matter and gauge field, respectively. 
Such a two-site two-particle model can be represented 
using the four basis states:
$\ketone\!=\!\ket{a,0}\otimes\ket{f,0}$, $\kettwo\!=\!\ket{a,0}\otimes\ket{0,f}$,
$\ketthree\!=\!\ket{0,a}\otimes\ket{f,0}$ and $\ketfour\!=\!\ket{0,a}\otimes\ket{0,f}$, 
where the labels $a$ and $f$ before or after the comma mark the
particle occupation on the left and right site.
As described in the main text,
a species-dependent energy offset $\Delta_f\!=\!U$ between the two sites
and a species-independent resonant driving $A \cos(\omega t + \phi)$ 
of the on-site potential are applied,
which leads to the time-dependent Hamiltonian~\eqref{eq:Ht} in the main text.
In the new basis defined above, this Hamiltonian reads
\begin{align}
\hat{H}(t) = 
    & - J \left(\ketthree\braone + \ketfour\bratwo +
          \ketfour\brathree + \kettwo\braone + \hc\right)\nonumber\\
    & + U \left(2\ketone\braone + \ketthree\brathree + 
          \ketfour\brafour\right)\label{eq:suppl:timedepH}\\
    & + A \cos(\omega t + \phi)\left(2\ketone\braone + 
          \kettwo\bratwo + \ketthree\brathree\right),\nonumber
\end{align}
where the tunneling rate of both $a$- and $f$-particles is $J$ 
and the intra-species interaction energy is $U$.

The stroboscopic dynamics of such a time-dependent system
can be described by an effective Floquet Hamiltonian
represented by a series of time-independent terms in powers of $1/\omega$.
The series can be truncated to lowest order
in the high-frequency limit $\omega\rightarrow\infty$.
Following the method presented in \cite{Goldman:2015kca:suppl, Goldman:2014:suppl},
we calculate the Floquet Hamiltonian $\hat{H}_\mathrm{F}$
up to first order.
To this end,
Eq.~\eqref{eq:suppl:timedepH} is transformed by a unitary transformation
$\ket{\psi}\rightarrow \ket{\psi^\prime}\!=\!\hat{R}(t)\ket{\psi}$,
such that the new Hamiltonian $\hat{\mathcal{H}}(t)$
does not contain divergent terms in the high-frequency limit:
\begin{equation}
\hat{\mathcal{H}}(t)
   = \hat{R}\hat{H}(t)\hat{R}^\dagger 
     - \ii \hat{R}\partial_{t} \hat{R}^\dagger
   = \sum\limits_{k\in\mathbb{Z}} \hat{\mathcal{H}}^{(k)} \e^{\ii k\omega t}.
\end{equation}
We express this transformed Hamiltonian in a Fourier-series
with time-independent components~$\hat{\mathcal{H}}^{(k)}$.
For our  model~\eqref{eq:suppl:timedepH},
we choose the transformation
\begin{align}
\hat{R}(t) = & \exp\Bigl\{
      + \ii\omega t \left(2\ketone\braone + \ketthree\brathree
                 + \ketfour\brafour\right)\\
     & + \frac{\ii A}{\hbar\omega}\sin(\omega t + \phi)
     		\left(2\ketone\braone + \kettwo\bratwo + 
     				\ketthree\brathree\right)
	\Bigr\}\nonumber
\end{align}
and obtain the following Fourier components
\begin{widetext}
\begin{align}
\hat{\mathcal{H}}^{(k)} =& -J\,\Bigl\{
      \besselchi{-k-1} \ketthree\braone\e^{\ii(k+1)\phi}
    + \besselchi{k-1}  \ketone\brathree\e^{\ii(k-1)\phi}
    + \besselchi{-k+1} \ketfour\bratwo\e^{\ii(k-1)\phi}
    + \besselchi{k+1}  \kettwo\brafour\e^{\ii(k+1)\phi} \nonumber\\&
    + \besselchi{-k}   \ketfour\brathree\e^{\ii k\phi} 
    + \besselchi{k}    \ketthree\brafour\e^{\ii k\phi}
    + \besselchi{-k-2}  \kettwo\braone\e^{\ii(k+2)\phi}
    + \besselchi{+k-2}\ketone\bratwo\e^{\ii(k-2)\phi}
      \Bigr\} \label{eq:suppl:effHk}\\&
    + (U-\hbar\omega)\delta_{k0}\,\Bigl\{2\ketone\braone 
    + \ketthree\brathree + \ketfour\brafour   
    \Bigr\}.\nonumber
\end{align}
\end{widetext}
Here $\besselchi{\nu}$ is the $\nu$-th Bessel function of the first kind 
with the dimensionless driving strength~$\chi\!=\!A/\hbar\omega$
and $\delta_{ij}$ the Kronecker delta.
Note that in the finite-frequency regime the resonant driving frequency
for a double well tilted by the energy difference~$U$
is equal to the energy gap, $\hbar\omega\!=\!\sqrt{U^2 + 4J^2}\neq U$.
In the high-frequency limit $U\gg J$, however, $\hbar\omega \simeq U$.
The lowest orders of the Floquet Hamiltonian can be calculated
using the components~$\hat{\mathcal{H}}^{(k)}$ [Eq.~\eqref{eq:suppl:effHk}]
according to
\begin{equation}
\hat{H}_\mathrm{F} = \hat{\mathcal{H}}^{(0)} 
    + \frac{1}{\hbar\omega}\sum\limits_{k>0}
    \frac{1}{k}\left[\hat{\mathcal{H}}^{(+k)}, \hat{\mathcal{H}}^{(-k)}\right]
    + \mathcal{O}\left(\frac{1}{\omega^2}\right).
    \label{eq:suppl:HeffDefinition}
\end{equation}

\subsection{Floquet model in the infinite-frequency limit}
In the limit $U/J \rightarrow\infty$,
the resonant driving frequency is  $\hbar\omega\!\simeq\!U$
and the diagonal terms in Eq.~\eqref{eq:suppl:effHk} vanish.
The Floquet Hamiltonian~\eqref{eq:suppl:HeffDefinition} reduces to the lowest order
\begin{equation}
\begin{aligned}
\hat{H}_\mathrm{F}^{0} =& -J \big\{
     \bessel[1]{\chi}\ketthree\braone \e^{\ii(\pi+\phi)}
    +\bessel[1]{\chi}\ketfour\bratwo \e^{-\ii\phi}\\
    &+\bessel[0]{\chi}\ketfour\brathree
    +\bessel[2]{\chi}\kettwo\braone \e^{2\phi}
    + \hc\bigr\},
\end{aligned}
\label{eq:suppl:effH0}
\end{equation}
referred to as zeroth-order Floquet Hamiltonian in the following.
As discussed in the main text, 
different tunnel processes have different renormalizations
with an additional phase factor depending on the modulation phase~$\phi$.
The tunnel processes
$\ketthree\braone$ and $\ketfour\bratwo$ correspond to $a$-particle tunneling
and $\ketfour\brathree$ and $\kettwo\braone$ to $f$-particle tunneling.
The $a$-particle's tunneling rate is renormalized 
with the first-order Bessel function of the first kind~$\besselchi{1}$,
while on the other hand the $f$-particle's tunneling rate
is renormalized depending on the $a$-particle's position.
The latter asymmetry
can be resolved by choosing the driving strength such 
that $\bessel[0]{\chi_0}\!=\!\bessel[2]{\chi_0}$.
This happens the first time at $\chi_0\approx1.84$ (Fig.~\ref{fig:2} in main text).
When choosing the modulation phase $\phi\!=\!\{0,\pi\}$,
the Floquet Hamiltonian~\eqref{eq:suppl:effH0} directly realizes 
the \Ztwo{} double well,
where the $a$-particle's tunnel phase is either $0$ or~$\pi$
depending on the $f$-particle's position, 
while tunneling of the $f$-particle is real-valued.
This density-dependent phase shift
reflects the sign of the effective energy offset~$\pm U$
between neighboring sites
experienced by the $a$-particle
because of the reflection properties of the Bessel function
$\besselchi{-1}\!=\!\e^{\ii\pi}\besselchi{1}$.

\subsection{Floquet model including the first-order correction for finite-frequency drive}
\begin{figure}
\includegraphics{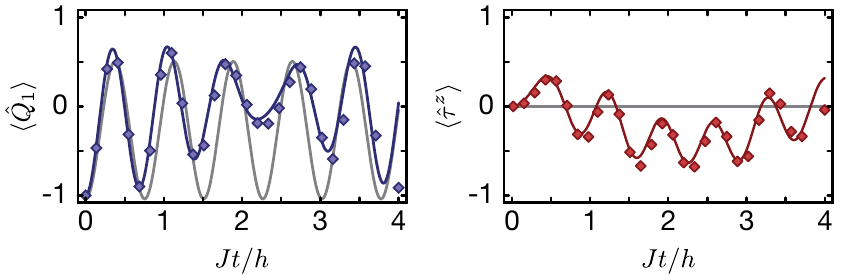}
\caption{\textbf{Finite-frequency corrections 
to the zeroth-order Floquet Hamiltonian.}
Solid lines are the time evolution
of the charge $\langle \hat{Q}_1 \rangle$ (blue)
and gauge field $\langle\hat{\tau}^z\rangle$ (red) 
according to \eqref{eq:suppl:timeevolveFloquet}
with the two lowest-order terms of the 
effective Floquet Hamiltonian~\eqref{eq:suppl:effHwithCorrections}.
Diamonds show the stroboscopic dynamics of the 
full time-dependent Hamiltonian~\eqref{eq:Ht} 
from the main text for resonant driving 
at $\hbar\omega\!=\!\sqrt{U^2+4J^2}\!\approx\! 1.04\,U$ and $U/J\!=\!7$.
Gray lines show the ideal solution Eq.~\eqref{eq:ideal}
of the zeroth-order effective Hamiltonian~\eqref{eq:suppl:effH0}
for $\phi=0$.
}
\label{fig:suppl:floquetcorrections}
\end{figure}

In experiments, it is generally challenging 
to work deep in the high-frequency limit,
especially without the availability of Feshbach resonances;
here, we work at $U/J\approx6.6$.
Consequently, higher order corrections become relevant.
In order to study their impact on the dynamics 
we calculate the first-order terms of the Floquet expansion 
and include them in our calculations (Fig.~\ref{fig:suppl:floquetcorrections}).
We restrict the derivation 
to stroboscopic time points $n\,T$
with $T\!=\!2\pi/\omega$, $n\in\mathbb{N}$
and set the driving phase to $\phi\!=\!0$.

The first-order term of the Floquet expansion 
can be calculated following Eq.~\eqref{eq:suppl:HeffDefinition}. 
Note, the resonant driving frequency is now $\hbar\omega\!=\!\sqrt{U^2 + 4J^2}\neq U$
and the diagonal terms in $\hat{\mathcal{H}}^{0}$ become nonzero.
\begin{widetext}
\begin{align}
\hat{H}_\mathrm{F} =
    &\begin{aligned}[t] \hat{\mathcal{H}}^{(0)}
    +\frac{J^2}{\hbar\omega}\sum\limits_{k>0}\frac{1}{k}
    &\Bigl\lbrace 
    \ketthree\bratwo\bigl(
      \besselchi{-k-1}\,\besselchi{-k-2} + \besselchi{k}\,\besselchi{k+1}
     -\besselchi{k-1}\,\besselchi{k-2} - \besselchi{-k}\,\besselchi{-k+1}
    \bigr)+\hc\\
   +& \ketfour\braone\bigl(
      \besselchi{-k+1}\,\besselchi{k-2} + \besselchi{-k}\,\besselchi{k-1}
     -\besselchi{k+1}\,\besselchi{-k-2} - \besselchi{+k}\,\besselchi{-k-1}
    \bigr)+\hc \Bigr\rbrace\\
  +\frac{J^2}{\hbar\omega}\sum\limits_{k>0}\frac{1}{k}
  &\Bigl\lbrace 
   \bigl(\ketone\braone - \kettwo\bratwo\bigr)
   \bigl(\besselchi{k-1}^2 + \besselchi{k-2}^2
      - \besselchi{-k-1}^2 - \besselchi{-k-2}^2
      \bigr)\\
   +&\bigl(\ketfour\brafour - \ketthree\brathree\bigr)
   \bigl(\besselchi{-k+1}^2 + \besselchi{-k}^2
      - \besselchi{k+1}^2 - \besselchi{k}^2\bigr)
    \Bigr\rbrace
    \end{aligned}\nonumber\\
    =&\hat{\mathcal{H}}^{(0)} + \hat{H}^{(1)}_\mathrm{pair-hopping} +
	\hat{H}^{(1)}_\mathrm{detuning}\label{eq:suppl:effHwithCorrections}
\end{align}
\end{widetext}
The first-order correction contains two types of terms:
a pair-tunneling term $\hat{H}^{(1)}_\mathrm{pair-hopping}$,
which corresponds to tunneling of both particles, 
and a detuning term $\hat{H}^{(1)}_\mathrm{detuning}$,
describing the deviation from the exact resonance condition.
 
When calculating the time evolution including higher order terms,
the time evolution operator~$\hat{U}$ gets modified in general by both 
the transformation $\hat{R}$ 
and the kick operator $\hat{K}$ \cite{Goldman:2015kca:suppl}.
However, for stroboscopic times the time-evolution operator reduces to
\begin{equation}
\hat{U}(0\rightarrow nT) = 
    e^{-\ii\hat{K}}e^{-\ii\hat{H}_\mathrm{F} nT/\hbar}e^{\ii\hat{K}}\label{eq:suppl:timeevolveFloquet}
\end{equation}
and only depends on the stroboscopic kick operator
\begin{equation}
\hat{K} = \frac{1}{\ii\hbar\omega}\sum\limits_{k>0}
    \frac{1}{k}\left(\hat{\mathcal{H}}^{(+k)} %e^{+\ii k \omega nT}
    - \hat{\mathcal{H}}^{(-k)} %e^{-\ii k \omega nT} 
    \right)
    + \mathcal{O}\left(\frac{1}{\omega^2}\right)
    .
\end{equation}

In Fig.~\ref{fig:suppl:floquetcorrections} the dynamics 
of the effective Floquet model including first-order corrections
is compared to a full time-dependent analysis 
of Hamiltonian~\eqref{eq:Ht} from the main text.
We find that for the experimental timescales and parameters, i.e. $U/J=7$,
the effective model up to first order
is sufficient to describe the dynamics of the system.

\section{Description of the experiment}
\subsection{Lattice setup}
\begin{figure}
\includegraphics{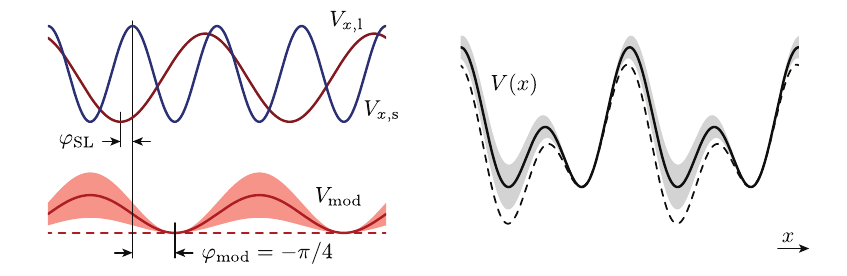}
\caption{\textbf{Lattice setup.}
A superlattice (black) is composed 
of a lattice with wavelength $\lambda_\mathrm{s}$ (blue)
and a lattice with wavelength $\lambda_\mathrm{l}\!=\!2\lambda_\mathrm{s}$ (red)
with a relative phase of $\varphi_\mathrm{SL}$.
In addition, a second lattice with wavelength $\lambda_\mathrm{l}$ is superimposed
with a relative phase of $\varphi_\mathrm{mod}\!=\!-\pi/4$, 
such that the on-site energy of the left double-well site 
can be modified by a change in the lattice depth~$V_\mathrm{mod}$ 
[Eq.~\eqref{eq:mod_lattice}]
illustrated by the red and gray shadings.
The dashed lines show the superlattice for  $V_\mathrm{mod}=0$.}
\label{fig:suppl:lattice}
\end{figure}

In the experiment, ultracold $^{87}$Rb atoms 
in a three-dimensional (3D) optical lattice potential are used.
The lattice consists of three mutually orthogonal standing waves 
with wavelength $\lambda_\mathrm{s}\!=\!767\,$nm 
and an additional lattice with 
$\lambda_\mathrm{l}\!=\!2\lambda_\mathrm{s}$ 
along the $x$-direction, 
which generates the superlattice potential 
\begin{equation}
V_\mathrm{SL}(x)\!=\!V_{x,\mathrm{s}} \cos^2(k_\mathrm{s} x) 
    + V_{x,\mathrm{l}} \cos^2(k_\mathrm{l} x
    + \varphi_\mathrm{SL}),
\end{equation}
where $V_{x,\mu}$ denotes the lattice depth 
and $k_{\mu}\!=\!2\pi/\lambda_{\mu}$ is the wave number,
with $\mu\in\{\mathrm{s},\mathrm{l}\}$.
In addition, a standing wave 
with wavelength $\lambda_\mathrm{l}$ is superimposed,
where the relative phase between the potentials is chosen
such that the potential maxima 
affect only one of the double-well sites
(Fig.~\ref{fig:suppl:lattice}).
The overall potential is then given by
\begin{equation}
V(x)\!=\!V_{\mathrm{SL}}(x) + V_{\mathrm{mod}}\,\cos^2(k_l x - \pi/4),
\label{eq:mod_lattice}
\end{equation}
with $V_{\mathrm{mod}}$ the lattice depth of the additional modulation lattice.
A time-dependent modulation $V_{\mathrm{mod}}%
\!=\!V_{\mathrm{mod}}^{(0)} + A_\mathrm{mod}\cos(\omega t + \phi)$
around a mean value $V_{\mathrm{mod}}^{(0)}$
in combination with a suitable static superlattice phase~$\varphi_\mathrm{SL}$
then generates the time-dependent Hamiltonian~\eqref{eq:suppl:timedepH}.

\subsection{Measurement of the renormalization of the tunnel coupling in a driven double well}
We experimentally determine the renormalization of the tunnel coupling 
of a single particle in a driven double well.
The time-dependent Hamiltonian can be written in the basis, where the particle occupies 
the left $\ket{L}$ or the right $\ket{R}$ site of the double well:
\begin{align}
\hat{H}_\mathrm{1P}(t) =& -J^\prime\left(\ket{L}\bra{R} + \ket{R}\bra{L}\right)\nonumber\\
    &+ \left\{\Delta+ A \cos(\omega t + \phi)\right\}\ket{L}\bra{L} 
    ,
\end{align}
where $J^\prime$ is the single-particle tunneling rate, 
$\Delta$ the energy offset between neighboring sites,
$A$ the modulation amplitude,
$\omega$ the modulation frequency
and $\phi$ the modulation phase.
In the high-frequency limit
with resonant multi-frequency driving $\Delta\!=\!\nu\,\hbar\omega$, with $\nu\in\mathbb{Z}$,
we find the Fourier components of the Floquet expansion 
analogous to Eq.~\eqref{eq:suppl:effHk}
\begin{align}
\hat{\mathcal{H}}^{(k)}_\mathrm{1P} = -J^\prime\Bigl\{
 &\bessel[-k-\nu]{\chi}\,\e^{\ii(k + \nu)\phi} \ket{R}\bra{L} + \nonumber\\
 &\bessel[+k-\nu]{\chi}\,\e^{\ii(k - \nu)\phi} \ket{L}\bra{R}
\Bigr\}.
\end{align}
In the infinite frequency limit $\omega\rightarrow\infty$, 
we can extract the renormalization of the tunnel coupling
$\widetilde{J}_\nu\!=\!J^\prime|\bessel[\nu]{\chi}|$, 
which scales with the $\nu$th-order Bessel function of the first kind.

We load atoms in the $\ket{F\!=\!1, m_F\!=\!-1}$ state
into a 3D optical lattice by performing an exponential-shaped ramp
during a time segment of $50\,$ms;
the final lattice depths $V_{x,\mathrm{l}}\!=\!30(1)\,E_{r,\mathrm{l}}$, 
$V_y\!=\!35(1)\,E_{r,\mathrm{s}}$ 
are reached within the $1/e$-time $\tau\!=\!5\,$ms, 
while the vertical lattice $V_z\!=\!50(1)\,E_{r,\mathrm{s}}$ 
is ramped up within $\tau_z\!=\!1.5\,$ms. 
The lattice depth is given in units of their respective recoil energy
$E_{r,\mu}\!=\!\hbar^2k_\mu^2/2m$ with $m$ the mass of a rubidium atom and $\mu\!=\!\{\mathrm{s}, \mathrm{l}\}$.
Then, we use a filtering sequence to remove doubly-occupied sites
by light assisted-collisions between atoms in the $|F\!=\!2, m_F\!=\!-1\rangle$ state.
The subsequent part of the sequence 
is in principle not needed for this measurement
but it is part of the final experimental sequence and it is kept 
in order to reach a comparable experimental situation.
In this part 
the long-period lattice site is split within $15\,$ms
by increasing $V_{x,\mathrm{s}}\!=\!30(1)\,E_{r,\mathrm{s}}$ 
together with the transversal lattices $V_{y,\mathrm{s}}\!=\!V_{z,\mathrm{s}}=60(2)\,E_{r,\mathrm{s}}$ 
and merged again in $15\,$ms by ramping down $V_{x,\mathrm{s}}\!=\!0$.
Then, the modulation lattice is turned on at the mean value
$V_{\mathrm{mod}}^{(0)}\!=\!15.0(2)\,E_{r,\mathrm{l}}$ in $20\,$ms
and the superlattice phase~$\varphi_\mathrm{SL}$ is tuned to a value 
such that $\Delta\gg J$.
The particle localizes to one site during the subsequent splitting process,
where the short lattice is ramped to 
$V_{x,\mathrm{s}}\!=\!40(1)\,E_{r,\mathrm{s}}$ in $10\,$ms.
Subsequently, $\varphi_\mathrm{SL}$ is non-adiabatically changed to reach 
the final energy offset between the two sites~$\Delta_\nu$ in $10\,$ms.
The time evolution begins with a rapid coupling of the two sites 
by decreasing $V_{x,\mathrm{s}}\!=\!9.5(1)\,E_{r,\mathrm{s}}$ in $100\,\mu$s
and starting the modulation with frequency $\omega\!=\!2\pi\times4122\,$Hz.
This procedure was repeated for various modulation amplitudes~$A_\mathrm{mod}$.
For the detection, we freeze the motion in $100\,\mu$s 
and use site-resolved band mapping to determine the site occupations
$N_\mathrm{L}$ and $N_\mathrm{R}$
from which we determine the site imbalance
$I\!=\!(N_\mathrm{L} - N_\mathrm{R})/(N_\mathrm{L} + N_\mathrm{R})$
\cite{SebbyStrabley:2006gj:suppl, Folling:2007jg:suppl}.

For finite driving frequencies,
corrections to the resonant driving condition 
have to be taken into account
$\nu\hbar\omega\!=\!\sqrt{\Delta_\nu^2 + 4J^2}$ with $\nu\neq0$.
Thus, we have experimentally determined $\Delta_\nu$ using a spectroscopic measurement 
by varying the energy offset at constant driving frequency~$\omega$.
Measurements of a time-trace for different modulation amplitudes~$A_\mathrm{mod}$
were performed for $\nu \in \{0, 1, 2\}$.
Each resulting imbalance time-trace represents an average 
over the time-traces of all double wells in a 3D array.
For technical reasons, each double well in this 3D array experiences
different energy offsets between neighboring sites.
To model this effect, we assume
that these energy offsets are distributed according to
\begin{equation}
\mathcal{G}_\Delta(\delta) = 
\frac{1}{\sqrt{2\pi\Delta_\sigma^2}}\,\exp\left(-\frac{\delta^2}{2\Delta_\sigma^2}\right),
\label{eq:suppl:GaussianDelta}
\end{equation}
with a standard deviation $\Delta_\sigma$.
Taking this effect into account,
the renormalized tunneling rates~$\widetilde{J}_\nu$ were extracted 
by fitting an average of $S\!=\!10$ sinusoidal functions
\begin{equation}
I_\nu(t) = \frac{A}{S}\sum\limits_{n\!=\!1}^S \sin\left(\sqrt{\delta_n^2 + 4\widetilde{J}_\nu^2}\,t/\hbar + \xi\right) + I_0
\end{equation}
to the data, where $A$ is the oscillation amplitude, 
$\delta_n$ a random sample of the tilt distribution~\eqref{eq:suppl:GaussianDelta},
$\widetilde{J}_\nu\!=\!J^\prime_\nu\besselchi{\nu}$ is the renormalized tunnel coupling, 
$\xi$ is an initial phase due to the finite initialization ramp time, 
and $I_0$ the imbalance offset of the oscillation.
The fit uses five free fit parameters:
$\widetilde{J}_\nu$, $A$, $\Delta_\sigma$, $\xi$, and $I_0$.
To estimate the confidence interval we performed a bootstrap analysis with $1000$~realizations
for different sets of $\{\delta_n\}$, 
a Gaussian error in the detection of the imbalance of $0.05$ 
and randomly-guessed fit start values for $\widetilde{J}_\nu$.
In Fig.~\ref{fig:2} (main text), 
$\widetilde{J}_\nu$ is normalized to its respective bare tunnel coupling strength~$J^\prime_\nu$
calculated from the double-well Wannier functions based on the calibrated lattice parameters:
$J^\prime_0/h\!=\!490\,$Hz, $J^\prime_1/h\!=\!520\,$Hz and $J^\prime_2/h\!=\!563\,$Hz.
They are slightly different for the three data sets, $\nu\!=\!\{0,1,2\}$.
The data points and error bars in Fig.~\ref{fig:2} show the median
and the $1\sigma$-confidence interval from the bootstrap analysis.
The conversion~$\alpha$ of the driving amplitude $A\!=\!\alpha\,A_\mathrm{mod}$ 
is calibrated by a fit of the function $\bessel[0]{\alpha_\mathrm{fit}\,A_\mathrm{mod}/\omega}$
to the experimental results for $\nu\!=\!0$.
A comparison of the fitted to the calculated value results in
$\alpha_\mathrm{fit}/\alpha_\mathrm{calc}\approx1.09(2)$.
The order of magnitude of this value is consistent for all measurements
and is most likely attributed to the non-linear behavior of $\Delta(V_\mathrm{mod})$.
In addition, also the tunneling rate 
depends on the value of the modulation lattice depth $J^\prime_\nu(V_\mathrm{mod})$
and changes therefore sightly during a driving period.
These effects are especially important
for large values $V_\mathrm{mod} \sim V_{x,\mathrm{l}}$,
because the modulation lattice changes not only the on-site energy, 
but also the combined lattice potential's shape~$V(x)$. 

\subsection{Species-dependent energy offset \\for the \Ztwo{} two-site model}
The species-dependent energy offset 
is realized by a combination of a magnetic gradient 
and a superlattice with a relative phase~$\varphi_\mathrm{SL}$.
The magnetic gradient 
induces opposite energy offsets~$\Delta_\mathrm{M}$
for the two hyperfine-states 
that encode the $a$- and $f$-particles, 
i.e. $\left|F\!=\!1, m_F\!=\!\mp1\right\rangle$,
because these states have opposite magnetic moments.
The experimental implementation 
requires an energy offset of zero between neighboring sites
for the $a$-particles. 
Thus, the species-independent energy offset 
from the superlattice potential 
is chosen, such that it compensates
the tilt $\Delta_\mathrm{M}\!=\!-\Delta_\mathrm{SL}$.
The energy offset between neighboring sites
for the $f$-particles is then 
$\Delta_f\!=\!\Delta_\mathrm{M}+\Delta_\mathrm{SL}$,
which needs to be matched with the interaction energy~$U$.

\begin{figure*}
\includegraphics{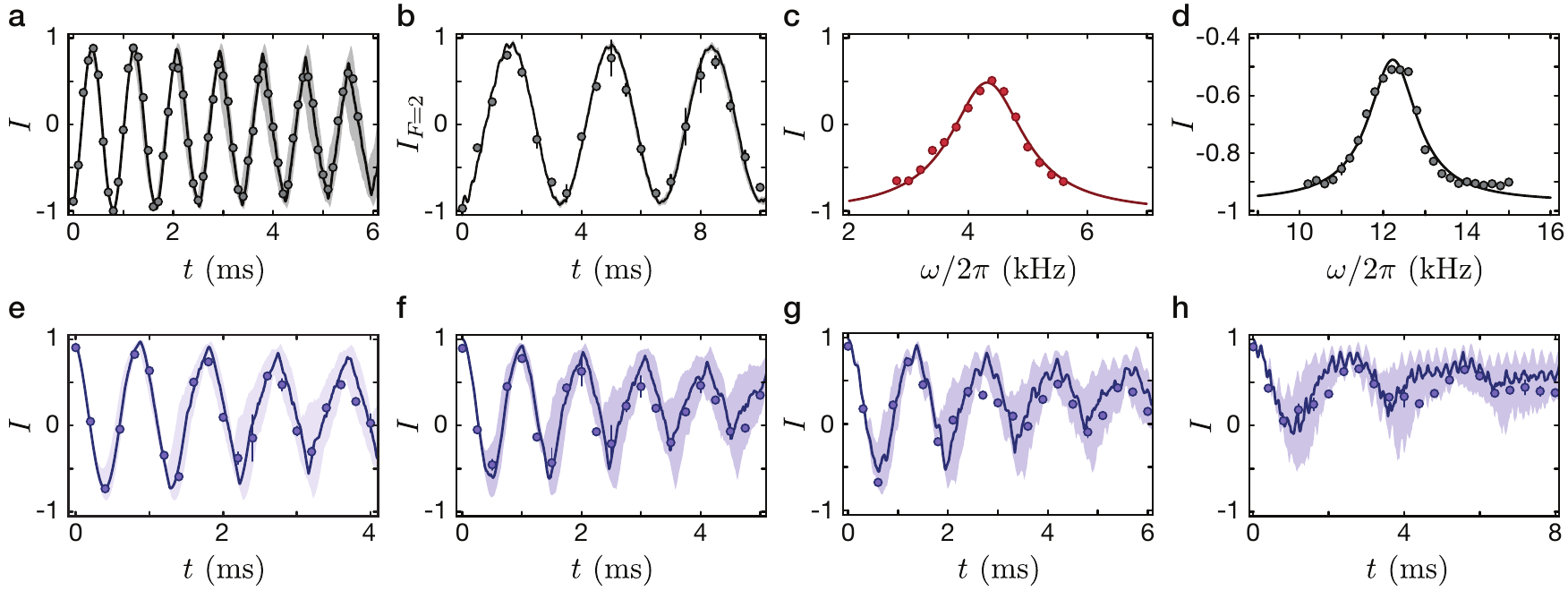}
\caption{\textbf{Parameter calibrations.}
\textbf{a} Tunnel oscillations in a symmetric double well
for $V_{x,\mathrm{s}}\!=\!9.50(1)\,E_{r,\mathrm{s}}$, $V_{x,\mathrm{l}}\!=\!35(1)\,E_{r,\mathrm{l}}$,
and $V_{\mathrm{mod}}\!=\!0$. 
The resulting single-particle tunneling rate is $J^\prime/h\!=\!574(3)\,$Hz.
\textbf{b} Superexchange oscillations.
\textbf{c} The spectroscopic measurement of the tilt~$\Delta_f\!=\!\Delta_\mathrm{M} + \Delta_\mathrm{SL}$
leads to the resonance $\omega\!=\!2\pi \times 4.32(2)\,$kHz.
\textbf{d} A spectroscopic measurement of the tilt induced only by the modulation lattice
$V^{(0)}_{\mathrm{mod}}\!=\!15.0(1)\,E_{r,\mathrm{l}}$
leads to a resonance $\omega_\mathrm{mod}\!=\!2\pi \times 12.23(2)\,$kHz.
\textbf{e} Tunnel oscillations in a symmetric configuration for the final lattice parameters
$V_{x,\mathrm{s}}\!=\!9.50(1)\,E_{r,\mathrm{s}}$, $V_{x,\mathrm{l}}\!=\!35(1)\,E_{r,\mathrm{l}}$,
and $V_{\mathrm{mod}}^{(0)}\!=\!15.0(2)\,E_{r,\mathrm{l}}$ in the presence of the magnetic gradient.
The particle experiences a potential
equivalent to the one seen by the $a$-particle.
The resulting tunneling rate is $J^\prime/h\!=\!523(3)\,$Hz.
\textbf{f-h} Driven tunnel oscillations in the same configuration as \textbf{e} 
for $\chi\approx0.61$ in \textbf{f}, $\chi\approx1.23$ in \textbf{g}, 
$\chi\approx1.84$ in \textbf{h}.}
\label{fig:suppl:calibrations}
\end{figure*}

\subsection{Tight-binding description of the two-site two-particle model}
The working principle of the experimental implementation for the \Ztwo{} two-site model
is discussed in detail in the main text.
Here, we present a tight-binding description 
including the full set of experimental parameters
and terms from the extended Bose-Hubbard model,
which are known to exist for interacting particles in lattices~\cite{Scarola:2005:suppl}.
The extended Bose-Hubbard model takes into account
that the Wannier functions~$w_\mathrm{\mu}(r)$
of particles on different sites~$\mu$ overlap.
Hence, this leads to nearest-neighbor interactions 
$u_\text{LR}\!=\!g\int |w_\mathrm{L}(r)|^2 |w_\mathrm{R}(r)|^2\,\mathrm{d}r^3$ 
and density-assisted tunneling,
where the tunnel matrix element $J\!=\!J^\prime + \delta j$ gets modified by
$\delta j\!=\!g\int w_\mathrm{L}^{*2}(r)w_\mathrm{L}(r)w_\mathrm{R}(r)\,\mathrm{d}r^3$.
In addition, two-particle hopping processes arise with a strength~$u_\text{LR}$,
where either two particles on the same site tunnel simultaneously to a neighboring site,
or two neighboring particles exchange their positions.
Note, the two-particle hopping processes directly break the \Ztwo{} symmetry
but for our parameters the value is small, $u_\mathrm{LR}/J \approx 0.03$,
compared to the experimental timescales.
All terms of the extended Bose-Hubbard model
are proportional to the effective interaction strength
$g\!=\!4\pi\hbar^2 a_\mathrm{s}/m$,
with $a_\mathrm{s}$ the inter-species s-wave scattering length
and $m$ the mass of a rubidium atom.
In summary, the time-dependent Hamiltonian~\eqref{eq:Ht}
generalizes to 
\begin{equation}
\begin{aligned}
H(t) =
    & - J  \bigl(\ketthree\braone + \ketfour\bratwo +
          \ketfour\brathree + \kettwo\braone \\&\qquad
        + \hc \bigl)\\
    & + U  \bigl(\ketone\braone + \ketfour\brafour \bigl)\\
    & + A \cos(\omega t) \bigl(\ketone\braone + 
          \kettwo\bratwo + \ketthree\brathree \bigl)\\
    & + \Delta_\mathrm{SL} \bigl(\ketone\braone + 
          \kettwo\bratwo + \ketthree\brathree \bigl)\\
    & + \Delta_\mathrm{M}  \bigl(\ketone\braone - 
          \kettwo\bratwo + \ketthree\brathree \bigl)\\
	& + u_\mathrm{LR} \bigl(\ketone\brafour + \kettwo\brathree + 
		  \ketfour\braone + \ketthree\bratwo \\&\qquad + 
          \kettwo\bratwo + \ketthree\brathree\bigr).
\end{aligned}
\label{eq:suppl:expHt}
\end{equation}

To realize the \Ztwo{} two-site model,
all parameters in Hamiltonian~\eqref{eq:suppl:expHt} need to be calibrated separately.
Especially, the magnetic gradient 
needs to be matched with the superlattice energy offset
$\Delta_\mathrm{M}\!=\!-\Delta_\mathrm{SL}$ 
to obtain zero tilt for the $a$-particles. 
This leads to a tilt $\Delta_f\!=\!\Delta_\mathrm{SL} + \Delta_\mathrm{M}$ 
for the $f$-particles,
which needs to be matched with the interaction energy $\Delta_f\!=\!U$.
Furthermore, we drive the system resonantly at
$\hbar\omega\!=\!\sqrt{\Delta_f^2 + 4J^2}$.
In the following we describe how the parameters are calibrated.

\subsection{Parameter calibrations}
\textit{Tunnel coupling in a symmetric double well} --- %
A single particle is localized on one site of a double well
with superlattice parameters $V_{x, \mathrm{l}}\!=\!35(1)\,E_{r,\mathrm{l}}$
and $V_{x, \mathrm{s}}\!=\!9.5(1)\,E_{r,\mathrm{s}}$.
From the measured tunnel oscillation frequency (Fig.~\ref{fig:suppl:calibrations}a)
we infer a precise value of the lattice depth $V_{x, \mathrm{s}}$.
The measurement sequence is analogous to the sequence 
for the renormalized tunnel coupling discussed before,
where we measure the site-imbalance at various times.
This time trace was then modeled using a two-site Bose-Hubbard Hamiltonian, 
including Gaussian distributed energy offsets~$\mathcal{G}_\Delta(\delta)$
[see Eq.~\eqref{eq:suppl:GaussianDelta}].
To this end, time traces for $256$ randomly-sampled values $\delta_n$ were calculated
and its median least-square fitted to the measured data set.
The results for the free parameters of this fit are 
$V_{x, \mathrm{s}}\!=\!9.50(1)\,E_{r,\mathrm{s}}$ and 
a tilt distribution with a standard deviation of~$\Delta_\sigma/h\!=\!0.44(4)\,$kHz.

\textit{On-site interaction energy} --- %
The on-site interaction energy $U$ was calibrated 
by a measurement of superexchange oscillations.
To this end, we localize two distinguishable particles
to the left and right site of a double well.
Then we couple the two sites and observe the oscillations of individual species.
For a known single-particle tunneling rate,
the interaction energy can be inferred from the measured oscillation frequency,
which scales as $J^{\prime2}/U$.
In the experiment, 
the on-site interaction energy depends strongly on the chosen lattice parameters,
thus a measurement in the final configuration is preferred.
Directly using $a$- and $f$-particles is not possible 
as they have opposite magnetic moments 
and the $f$-particle experiences an energy offset between neighboring sites.
However, we can use a microwave-driven adiabatic passage to transfer the 
$f$-particle to the state $\ket{F\!=\!2,\,m_F\!=\!+1}$ in the $F\!=\!2$ manifold,
such that it has the same magnetic moment as the $a$-particle.
In this configuration we can measure site-imbalance traces 
of the superexchange oscillations by $F$-state selective imaging
(Fig.~\ref{fig:suppl:calibrations}b) 
and calibrate the interaction energy $U=3.85(7)\,$kHz.

\textit{Magnetic and superlattice tilts} --- %
The energy offset between neighboring sites of the double well 
is given by a combination of the magnetic gradient,
the tilted superlattice potential and the modulation lattice.
The modulation lattice is static at $V_{\mathrm{mod}}^{(0)}\!=\!15.0(2)\,E_{r,\mathrm{l}}$
and the resulting tilt can be fully compensated by the superlattice phase
(Fig.~\ref{fig:suppl:lattice}).
Starting form this situation, we introduce a magnetic gradient 
and measure the imbalance~$I_a$ of a single $a$-particle.
At the superlattice phase where $I_a\!=\!0$
the tilts are identical $\Delta_\mathrm{SL}\!=\!\Delta_\mathrm{M}$.
Then we measure the energy offset $\Delta_f\!=\!\Delta_\mathrm{SL} + \Delta_\mathrm{M}$
introduced for the $f$-particles by modulation spectroscopy with the modulation lattice
(Fig.~\ref{fig:suppl:calibrations}c).
From the measurement we fit the resonance frequency~$\omega$.
This procedure was iteratively applied until the condition
$U\!=\!\Delta\!=\!\sqrt{\hbar^2\omega^2 - 4J^2}$ was fulfilled.

\textit{Characterization of the modulation amplitude} --- %
First, $V_\mathrm{mod}$ was determined
by a spectroscopic measurement
of the energy difference between neighboring sites
induced exclusively by the modulation lattice.
Therefore, a single particle was loaded to the lower site of a double well 
with $\varphi_\mathrm{SL}\!=\!0$ and $V_\mathrm{mod}\approx 15\,E_{r, \mathrm{l}}$
at a fixed modulation lattice phase of $-\pi/4$ (Fig.~\ref{fig:suppl:calibrations}d).
From this outcome a first estimate of the modulation amplitude can be made,
however, more accurate results can be gained by determining
the oscillation frequency of driven tunnel oscillations.
We choose $\nu\!=\!0$
as in this case the driving is always resonant
and the observable depends strongly on the driving strength.
We start with zero driving $A\!=\!0$
and verify that the oscillation frequency
in the combined potential agrees with the theoretical expectations
(Fig.~\ref{fig:suppl:calibrations}e).
Then, we measure a set of driven tunnel oscillations at $A\neq0$
(Fig.~\ref{fig:suppl:calibrations}f-h).
The driving amplitude was then fitted 
to match the measured oscillation frequencies.
In this model we also take inhomogeneous energy offsets into account,
which are modeled to be Gaussian distributed [Eq.~\eqref{eq:suppl:GaussianDelta}].

\textit{Initial state preparation} --- %
As discussed in the main text,
we first prepare an initial state
$\left| \psi_0^x\right>\!=\!(\ketone + \kettwo)/\sqrt{2}%
\!=\!|a, 0\rangle \otimes \left(|f, 0 \rangle + 
| 0,f\rangle\right)/\sqrt{2}$,
where the $f$-particle is in an eigenstate of the electric field operator~$\hat{\tau}^x$ 
and the $a$-particle is localized to the left site of the double well.
This initial state probes a single subsector of the model.
Experimentally we achieve this 
by realizing the groundstate of a specially-designed static model.
To this end, we start with two distinguishable atoms $a$ and $f$
on a single long period lattice site
(sequence of the \Ztwo{} double-well model, see below).
Then, the magnetic gradient and the superlattice phase 
are adjusted according to the calibrations above.
We set an initial value of the modulation lattice 
and create the two sites by adiabatically
increasing the short lattice to $V_{x,\mathrm{s}}\!=\!9.5E_{r,\mathrm{s}}$.
The initial value of the modulation lattice is chosen such
that the imbalance of the $f$-particles is zero.
The prepared initial state is the groundstate of the system in this configuration
$\left|\psi^x_\mathrm{init}\right>\approx 
0.70\ketone + 0.71\kettwo + 0.087\ketthree + 0.036\ketfour$.

Furthermore, we prepare the localized initial state
$\left| \psi_0^z\right>\!=\!\kettwo\!=\!|a, 0\rangle \otimes |0, f \rangle$.
Therefore, we instead use $V_\mathrm{mod}^{(0)}$ as initial value for the modulation lattice.
The corresponding experimental initial state is
$\left|\psi^z_\mathrm{init}\right>\approx 
0.071\ketone + 0.98\kettwo + 0.025\ketthree + 0.15\ketfour$.

\begin{figure}
\includegraphics{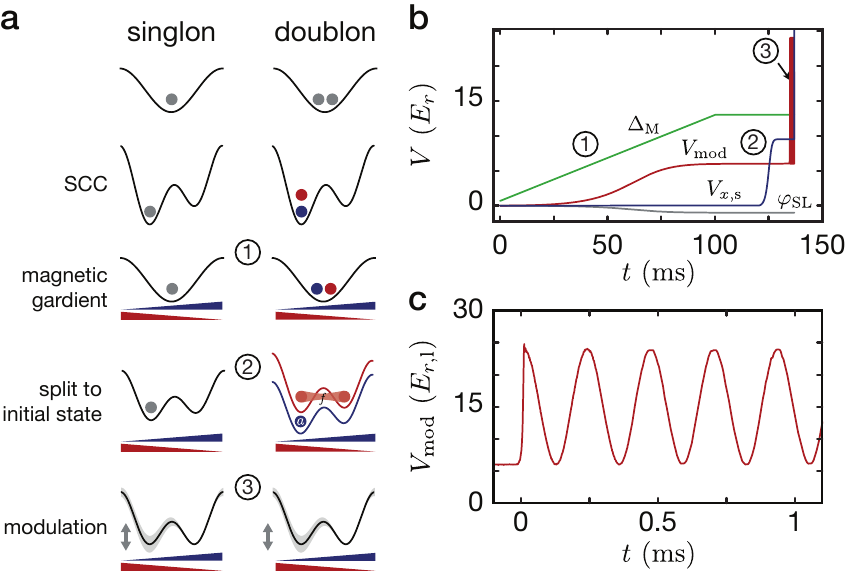}
\caption{\textbf{Experimental sequence.}
\textbf{a} Separate illustrations of the individual experimental steps 
for singly- and doubly-occupied double wells.
The dots denote atoms in the $F\!=\!1$ manifold of $^{87}$Rb
in the $m_F\!=\!-1$ (blue), $m_F\!=\!0$ (gray) and $m_F\!=\!+1$ (red) state.
The red and blue triangles illustrate the respective potential 
induced by the magnetic gradient.
\textbf{b} Parameter ramps for the numbered steps in \textbf{a}.
\textbf{c} Measured modulation lattice intensity,
with sudden jump at $t\!=\!0$.}
\label{fig:suppl:seq}
\end{figure}

\subsection{Experimental sequence for the \Ztwo{} two-site model}
The experimental sequence for the data presented in Fig.~\ref{fig:3} and~\ref{fig:4}
in the main text starts by loading atoms
in the $\ket{F\!=\!1, m_F\!=\!-1}$ state
into a 3D optical lattice
with lattice depths
$V_{x,\mathrm{l}}\!=\!30(1)\,E_{r,\mathrm{l}}$, 
$V_y\!=\!35(1)\,E_{r,\mathrm{s}}$ and $V_z\!=\!50(2)\,E_{r,\mathrm{s}}$.
These initial ramps are exponentially shaped 
and carried out in a time-segment of $50\,\textrm{ms}$ length.
The $1/e$-time of the exponential ramp
is $\tau\!=\!5\,$ms for the horizontal lattices ($x$, $y$)
and $\tau_z\!=\!1.5\,$ms for the vertical lattice ($z$).
This loading procedure was optimized 
for a maximal amount of doubly-occupied sites,
while loading a negligible amount of tripply- 
and a small amount of singly-occupied sites.
Later, each doubly-occupied site will realize 
a \Ztwo{} two-site model with both an $a$- and an $f$-particle.
To create these distinguishable atoms,
we use coherent spin-changing collisions (SCC)~\cite{Widera:2005:suppl}
to transfer atom pairs to the $\ket{F\!=\!1, m_F\!=\!\pm1}$ states
(Fig.~\ref{fig:suppl:seq}a).
To this end, we first apply
a series of microwave-driven adiabatic passages 
to transfer the atoms to the $\ket{F\!=\!1, m_F\!=\!0}$ state.
Subsequently, the short lattice in $x$-direction 
was ramped up to $V_{x,\mathrm{s}}\!=\!40(1)\,E_{r,\mathrm{s}}$ 
within $15\,\textrm{ms}$ 
with a superlattice phase $\varphi_{\mathrm{SL}}$,
such that both particles are localized to a single site of the double well.
At the same time the lattice depths along the orthogonal directions
are raised to $V_y\!=\!100(5)\,E_{r,\mathrm{s}}$ and $V_z\!=\!120(6)\,E_{r,\mathrm{s}}$
to further increase the on-site interaction energy.
Finally, an adiabatic passage of 
microwave-mediated SCC was performed within $100\,$ms.
Note, single atoms on a double well are not affected by SCC
and remain in the $\ket{F\!=\!1, m_F\!=\!0}$ state, 
which allows us to independently detect singly- and doubly-occupied sites.
After the SCC sequence, 
the double-well sites are merged again in $15\,\textrm{ms}$
by adiabatically switching off $V_{x, \mathrm{s}}\!=\!0$.
Then, a magnetic field gradient of 
$B^\prime/h\approx 5.9(3)\,\mathrm{kHz}/\lambda_\mathrm{s} $
was applied within $120\,\textrm{ms}$. 
Simultaneously, the modulation lattice is turned on 
to $V_{\mathrm{mod}}\!=\!6.0(1)\,E_{r,\mathrm{l}}$
and the superlattice phase ramped up to the final value.
The initial state is prepared by ramping up 
$V_{x,\mathrm{s}}\!=\!9.50(1)\,E_{r,\mathrm{s}}$ in $10\,$ms as described above.
Then, the modulation is started
with a sudden jump ($50\,\upmu$s) of the modulation intensity 
to obtain the correct initial driving phase of $\phi\!=\!0$ (Fig.~\ref{fig:suppl:seq}c). 
The modulation is performed with an amplitude of $A_\mathrm{mod}\!=\!9.0(1)\,E_{r,\mathrm{l}}$ 
around a mean value of $V_{\mathrm{mod}}^{(0)}\!=\!15.0(1)\,E_{r,\mathrm{l}}$
and a driving frequency of $\omega\!=\!2\pi\times 4320\,$Hz.
The initial state evolves in this driven model,
for different times~$t$,
approximating the time-dynamics in the \Ztwo{} two-site model.
Subsequently, the tunneling dynamics was inhibited 
by rapidly increasing the potential barrier 
between the two sites $V_{x,\mathrm{s}}\!=\!40(1)\,E_{r,\mathrm{s}}$ in $50\,\upmu$s,
followed by a site-resolved band-mapping detection
in combination with a Stern-Gerlach species separation~\cite{SebbyStrabley:2006gj:suppl,Folling:2007jg:suppl,Trotzky:2008jy:suppl}.

\begin{figure}
\includegraphics{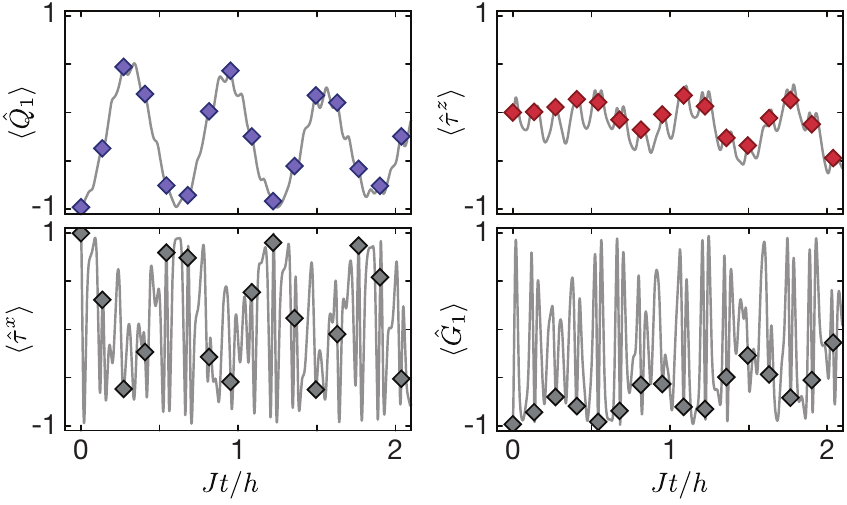}
\caption{\textbf{Time evolution for a single \Ztwo{} two-site model.}
Solid lines are the expectation values for the time evolution of
the \Ztwo{} charge $\langle\hat{Q}_1\rangle$,
the \Ztwo{} gauge field $\langle\hat{\tau}^z\rangle$,
the \Ztwo{} electric field $\langle\hat{\tau}^x\rangle$,
and the \Ztwo{} symmetry operator $\langle\hat{G}_1\rangle$.
The parameters are identical to the ones used in Fig.~\ref{fig:4}.
Note, the traces are for a single double well
and not averaged over a distribution of tilts.
Diamonds are stroboscopic time points.
}\label{fig:suppl:observablessingledoublewell}
\end{figure}

\section{Numeric time evolution}
The analysis of the effective Floquet model is compared to
a full, time-dependent numerical analysis of the two-site two-particle model.
We use formulation~\eqref{eq:suppl:expHt}, 
which is very close to the experimental realization
including the superlattice, magnetic gradient
and first-order corrections to the Bose-Hubbard model.

We use the experimentally calibrated parameters
$J^\prime$, $U$, $\Delta_\mathrm{M}$, $\Delta_\mathrm{SL}$, and $\omega$
together with calculated extended Bose-Hubbard parameters $\delta j$ and $u_\mathrm{LR}$
to perform a numerical time evolution.
The time evolution is performed using the Trotter method
by applying the quasi-static time evolution operator $\hat{U}_n\!=\!\exp\{-\ii\,H(t_n)\Delta t/\hbar\}$
for each time point $t_n\!=\!n\,\Delta t$ with $n\in\mathbb{N}$.
The time step $\Delta t\!=\!2\pi/(s\omega)$ is chosen 
to subsample the driving frequency with $s\sim50$.
For the initial state $\left|\psi^x_\mathrm{init}\right>$,
the groundstate of the configuration for the initial state preparation is chosen
as explained above.

A single calculated time-trace shows strong oscillatory behavior
(Fig.~\ref{fig:suppl:observablessingledoublewell}),
which is not present in the stroboscopic calculation of the Floquet Hamiltonian
and can be attributed to the micromotion.
The micromotion is an additional dynamics with a period equal to the driving frequency.

The experimentally measured observables are averaged over many realizations of double wells.
Due to inhomogeneities in the energy offsets between neighboring sites, 
the traces of individual double wells are different.
We model these inhomogeneties with a Gaussian distribution of tilts
as described in Eq.~\eqref{eq:suppl:GaussianDelta} 
with a standard deviation of $\Delta_\sigma$, 
extracted from the measurement of tunnel oscillations.
From this distribution we randomly draw $1000$ tilt values
and average the observables.

\begin{figure}
\includegraphics{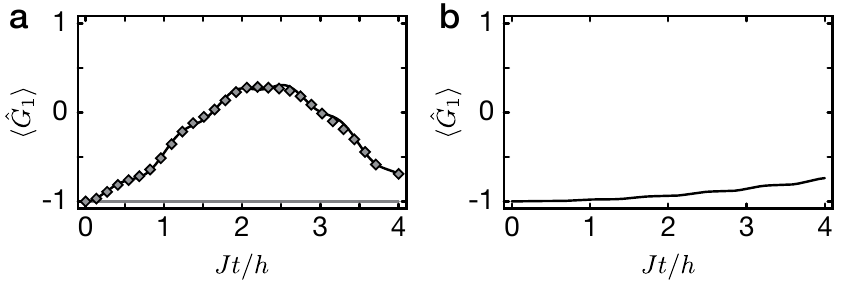}
\caption{\textbf{Symmetry breaking due to finite-frequency 
and extended Bose-Hubbard corrections.}
\textbf{a} The solid line is the time evolution 
according to the time-evolution operator~\eqref{eq:suppl:timeevolveFloquet} of the expectation value 
of the \Ztwo{} symmetry operator~$\langle\hat{G}_1\rangle$  
including the two lowest-order terms of the 
effective Floquet Hamiltonian~\eqref{eq:suppl:effHwithCorrections}.
Diamonds show the stroboscopic dynamics of the 
full time-dependent Hamiltonian~\eqref{eq:Ht} 
from the main text for resonant driving 
at $\hbar\omega\!=\!\sqrt{U^2+4J^2}\!\approx\!1.04\,U$ and $U/J\!=\!7$.
The gray line shows the ideal solution Eq.~\eqref{eq:ideal}.
The corresponding dynamics of $\langle \hat{Q}_1 \rangle$ 
and $\langle\hat{\tau}^z\rangle$ are shown in Fig.~\ref{fig:suppl:floquetcorrections}.
\textbf{b} Time evolution of the expectation value 
of the \Ztwo{} symmetry operator~$\langle\hat{G}_1\rangle$  
including additional terms from the extended Bose-Hubbard model
with a strength~$u_\mathrm{LR}\!=\!0.03\,J$
to the effective Floquet model in the infinite-frequency limit.
For all calculations the initial state was 
$\ket{\psi^x_\mathrm{init}}\!=\!(\ketone+\kettwo)/\sqrt{2}$.
}
\label{fig:suppl:symmetrybreaking}
\end{figure}

\section{Symmetry-breaking correction terms in the experimental realization}
It is essential for experimental realizations of LGTs
that the underlying gauge symmetry is sufficiently conserved during the experiment.
This puts high requirements on experimental implementations 
to suppress symmetry-breaking terms.
In the presented scheme, we find two main sources for symmetry breaking:
finite-frequency corrections to the effective Floquet Hamiltonian
and first-order corrections to the Bose-Hubbard model. 

The first-order correction to the Floquet Hamiltonian for finite-frequency drive
was derived above [Eq.~\eqref{eq:suppl:effHwithCorrections}].
The corrections contain diagonal detuning, 
correlated hopping and direct exchange coupling terms,
which scale as $J^2/(\hbar\omega)$.
The driving frequency is connected to the interaction energy in this realization.
Therefore, the corrections can be reduced by either increasing the interaction energy
or decreasing the tunnel coupling~$J$.
The effect on the \Ztwo{} symmetry is studied in Fig.~\ref{fig:suppl:symmetrybreaking}a
for $U/J\!=\!7$, similar to the experimental parameters,
and an initial state in the $g_1\!=\!-1$ subsector.

The additional terms from the extended Bose-Hubbard model
were introduced in Eq.~\eqref{eq:suppl:expHt}
and are nearest-neighbor interactions, correlated tunneling 
and a direct exchange coupling.
The terms scale with the square overlap of neighboring Wannier functions,
$u_\text{LR}\!=\!g\int |w_\mathrm{L}(r)|^2 |w_\mathrm{R}(r)|^2\,\mathrm{d}r^3$,
and vanish therefore faster than the tunnel coupling~$J$,
when reducing the Wannier overlap.
In Fig.~\ref{fig:suppl:symmetrybreaking}b
the correction terms were added to the infinite-frequency Floquet model
with strength $u_\mathrm{LR}\!=\!0.03\,J$
and the time evolution of the expectation value of~$\hat{G}$ was calculated numerically.

\begin{figure}
\includegraphics{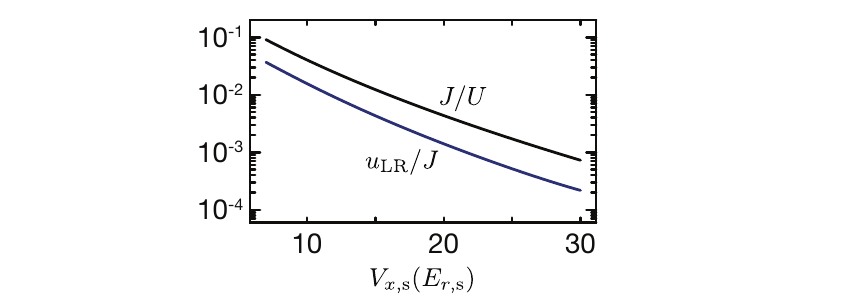}
\caption{\textbf{Suppressing the corrections.}
Scaling of the tight-binding parameters $J/U$ and $u_{\mathrm{LR}}/J$ with $V_{x,\mathrm{s}}$ and 
$V_{x,\mathrm{l}} \sim \sqrt{V_{x,\mathrm{s}}}$ for an initial ratio of
$V_{x,\mathrm{s}}/V_{x,\mathrm{l}} = 9.5/35$.
This relative scaling keeps the ratio of the inter- 
to the intra-double well tunneling constant.
}
\label{fig:suppl:scalingOfCorrections}
\end{figure}

Both symmetry-breaking terms can be reduced by increasing the short period lattice potential depth,
as the Wannier function overlap between neighboring sites reduces,
while the on-site overlap and therefore the interaction energy increases.
In Fig.~\ref{fig:suppl:scalingOfCorrections}, the relative scaling of the correction terms 
with~$V_{x,\mathrm{s}}$ are shown. 
Note, to keep the ratio between the inter-double well tunneling rate~$J_i$ 
and the tunneling rate~$J$ constant for comparability, 
the long period lattice depth was scaled accordingly 
with $V_{x,\mathrm{l}}\propto\sqrt{V_{x,\mathrm{s}}}$.

\begin{figure}
\includegraphics{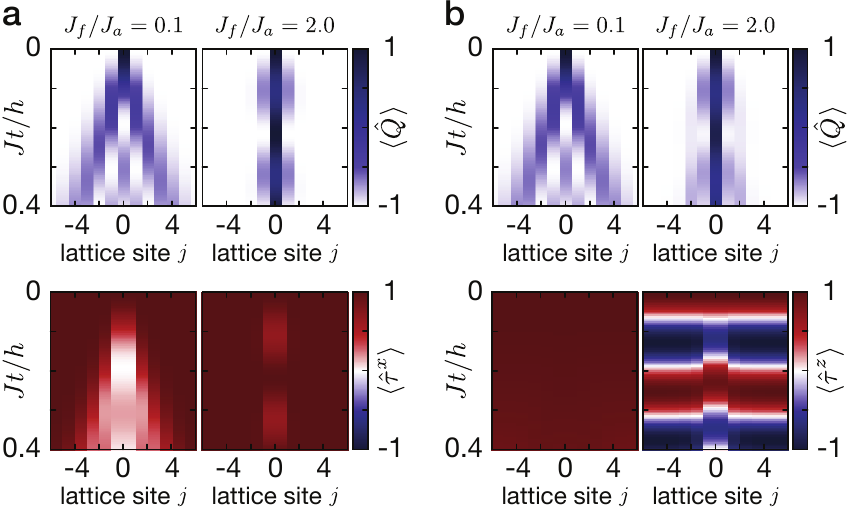}
\caption{\textbf{Numerical time evolution of a 1D $\mathbb{Z}_2$ chain 
with one matter particle.}
Dynamics of the matter particle and \Ztwo{} degrees-of-freedom
for two different initial states 
calculated using exact diagonalization 
on a system with 13 sites based on Eq.~\eqref{eq:H1D}. 
The matter particle is initially localized on site $j\!=\!0$.
\textbf{a} All links are initialized in $\tau^x_{\langle j,j+1\rangle}\!=\!+1$.
\textbf{b} All links are initialized in $\tau^z_{\langle j,j+1\rangle}\!=\!+1$.
}
\label{fig:suppl:1D}
\end{figure}

\section{\Ztwo{} LGT in 1D}
\subsection{The ideal 1D model}
The \Ztwo{} double well is a minimal two-site model,
which consists of one link and a single matter particle.
A direct generalization of this model arises 
when extending the double well to a chain by adding more links, 
an experimental scheme to realize such a system is introduced below.
We discuss the dynamics for the resulting \Ztwo{} chain 
described by Eq.~\eqref{eq:H1D} in the main text
in different regimes $J_f/J_a$ 
and for two initial states analogous to the ones used in the main text
to study the two-site model.
The dynamics of the system is analyzed 
by a calculation of the numerical time evolution with a Hamiltonian for $13$ sites.
Initially, the matter particle is always in the center of the system labeled~$j\!=\!0$.

We first start with an initial state,
where all links are in the groundstate 
of the electric field operator $-\hat{\tau}^x_{\langle j,j+1\rangle}$.
Therefore $\tau^x_{\langle j,j+1\rangle}\!=\!+1$ for all sites~$j$.
Together with the matter particle at site $j\!=\!0$,
this leads to $g_j\!=\!+1$ for all $j\neq0$ 
and a local static charge at $j\!=\!0$ described by $g_0\!=\!-1$.
As $\hat{G}$ commutes with the Hamiltonian~\eqref{eq:H1D},
the static charges~$g_j$ are conserved.
Thus, when the matter particle traverses a link
its electric field value needs to change.
This process is associated with an energy cost proportional to~$J_f$.
Figure~\ref{fig:suppl:1D}a summarizes the result for two different regimes.
In the regime $J_f\ll J_a$, 
the energy cost of flipping a traversed link, changing the value of $\tau^x$, is small
and the matter particle can freely expand, 
which is clearly visible in the expectation value of $\hat{Q}_j$.
The same cone shape is visible in the expectation value 
of the electric field operator $\langle\hat{\tau}^x\rangle$. 
Inside the cone $\langle\hat{\tau}^x\rangle\approx0$
because the particle either traversed the link or not.
In the opposite regime, where $J_f>J_a$, 
the energy cost for flipping $\tau^x$ is high, 
and also linearly increases with the number of traversed links.
Therefore, the matter particle stays confined to the local static charge at site $j\!=\!0$.

Analog to the experiment on the two-site model,
an initial state is examined,
where each link is in an eigenstate of the gauge field operator~$\hat{\tau}^z$.
In this situation,
instead of a single subsector
a superposition of many subsectors is probed.
We again investigate the two different regimes 
and present the results in Fig.~\ref{fig:suppl:1D}b.
In the regime $J_f \ll J_a$, 
the matter particle still expands freely,
while we again observe confinement of the matter particle 
in the regime $J_f > J_a$.
The expectation value of the electric field operator $\langle\hat{\tau}^x\rangle\!=\!0$
for all sites and times.
The expectation value of the gauge field operator $\langle\hat{\tau}^z\rangle$,
on the other hand, shows dynamics.
This dynamics can be explained on the level of a single link
prepared in one of the eigenstates of $\hat{\tau}^z$.
The coupling $-J_f\hat{\tau}^x$
leads to Rabi oscillations between $\tau^z\!=\!\pm1$.
In the presence of the matter particle 
the Rabi oscillations are detuned,
which leads to faster oscillations with a reduced amplitude.

\subsection{1D model with super-sites}
\textit{Super-site Hamiltonian} --- %
The \Ztwo{} double-well model is based on inter-species on-site interactions 
between $f$- and $a$-particles.
Simply connecting two double-wells by introducing a joint site 
shared by both leads to an ambiguous situation:
The $a$-particle can then no longer distinguish to which link the $f$-particles belong, 
unless different species $f$ and $f'$ are used on alternating links.

To resolve this complication, 
the \Ztwo{} double wells can be connected via a tunnel coupling~$J_c$ 
of the $a$-particle from the right site of the left double-well ($2j$) 
to the left site of the right double-well ($2 j +1$). 
Hence, the individual building blocks remain functional. 
The resulting chain of coupled double-well building blocks
can then be understood as a 1D model of super-sites at positions~$j$. 
Each super-site consists of matter creation operators $\hat{a}_j^\dagger$ and $\hat{b}_j^\dagger$,
describing the $a$-particles on sites $2j$ and $2j+1$, respectively. 
In this language, the super-sites are connected by a link 
with a gauge field~$\hat{\tau}^z_{\langle j, j\pm1\rangle}$.
The corresponding Hamiltonian is 
\begin{align}
\hat{H}^{\s}_{\mathrm{1D}} =& \sum\limits_j \left[
   - J^{\s}_a\,\hat{a}^\dagger_{j+1}\hat{\tau}^z_{\langle j,j+1\rangle}\hat{b}^{\phantom\dagger}_j
  - J^{\s}_c\,\hat{a}^{\dagger}_{j}\hat{b}^{\phantom\dagger}_{j} + \hc\right] \nonumber\\
  &- \sum\limits_j J^{\s}_f\,\hat{\tau}^x_{\langle j, j+1\rangle}.
\label{eq:suppl:H1Dsupersite}
\end{align}
When replacing the \Ztwo{} charge by the super-site charge
\begin{equation}
\hat{Q}^{\s}_{j} = \mathrm{e}^{\ii\pi(\hat{n}^a_j + \hat{n}^b_j)},
\label{eq:suppl:Z2ChargeSupersite}
\end{equation}
which depends on the matter particle occupation on the super-site $\hat{n}^s_j = \hat{n}^a_j + \hat{n}^b_j$,
then $\hat{H}_{\mathrm{1D},\,ab}$ is \Ztwo{} gauge-invariant with respect to the super-sites
and commutes with the gauge transformation 
$[\hat{H}^{\s}_{\mathrm{1D}},\hat{G}^{\s}_{jb}] = 0$ for all $j$.

\begin{figure}
\includegraphics{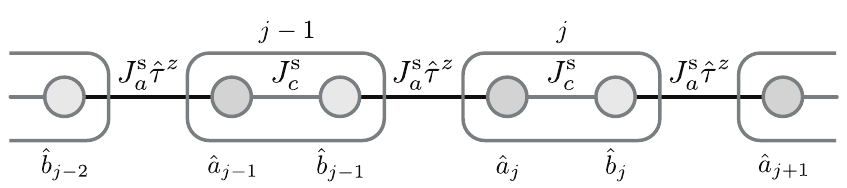}
\caption{\textbf{Illustration of the super-site model.}
Each site in the 1D \Ztwo{} chain is replaced by a super-site
with two physical sites~$a$ and~$b$.
They are coupled by a tunneling rate~$J^{\s}_\mathrm{c}$.
The links between the super-sites have a \Ztwo{} degree-of-freedom
and are realized by the building blocks.
}
\label{fig:suppl:IllustrationSupersiteModel}
\end{figure}

\textit{Ideal model limit of the super-site Hamiltonian} --- %
The super-site Hamiltonian reduces to the 1D model Eq.~\eqref{eq:H1D} in the main text, 
if $J^{\s}_c \gg J^{\s}_{a}$. 
In this limit, each super-site forms 
an energetically lower (upper) orbital $\hat{h}_{j,\pm} = (\hat{a}_j \pm \hat{b}_j)/\sqrt{2}$ 
with on-site energies described by 
$-J^{\s}_c (\hat{h}_{j,+}^\dagger \hat{h}_{j,+} - \hat{h}_{j,-}^\dagger \hat{h}_{j,-} ) = 
-J^{\s}_c (\hat{n}^h_{j,+} - \hat{n}^h_{j,-})$. 
Couplings between those higher and lower bands can be neglected 
in this limit $J^{\s}_{a} \ll J^{\s}_{c}$ 
and the effective Hamiltonian becomes
\begin{align}
\hat{H}^{\s}_{\mathrm{1D}} =& -J^{\s\star}_a \sum\limits_j \sum_{\mu = \pm} \mu\left[
    \,\hat{h}^\dagger_{j+1,\mu}\hat{\tau}^z_{\langle j,j+1\rangle}\hat{h}^{\phantom\dagger}_{j,\mu}  + \hc\right] \nonumber\\
  &- J^{\s}_c \sum\limits_j \left( \hat{n}^h_{j,+} - \hat{n}^h_{j,-} \right)\nonumber\\
  &- J^{\s}_f\sum\limits_j \hat{\tau}^x_{\langle j, j+1\rangle},
\label{eq:suppl:H1DsupersiteSimplified}
\end{align}
where $J^{\s\star}_a = J^{\s}_a/2$. 
At low energies, or in properly prepared states, the higher band remains unoccupied, 
$\hat{n}^h_{j,-} = 0$ for all $j$ and as a function of time. 
This yields a 1D \Ztwo{} LGT as discussed in the main text,
\begin{align}
\hat{H}^{\s}_{\mathrm{1D}} =&- J^{\s\star}_a \sum\limits_j  \left[
    \,\hat{h}^\dagger_{j+1,+}\hat{\tau}^z_{\langle j,j+1\rangle}\hat{h}^{\phantom\dagger}_{j,+}  + \hc\right] \nonumber\\
  &- J^{\s}_c \sum\limits_j  \hat{n}^h_{j,+}  - J^{\s}_f\sum\limits_j \hat{\tau}^x_{\langle j, j+1\rangle}.
\label{eq:suppl:H1DsupersiteSimplifiedFull}
\end{align}

If Hubbard interactions $ \hat{n}^a_j (\hat{n}^a_j-1) U_a/2$ 
and $ \hat{n}^b_j (\hat{n}^b_j-1) U_b/2$ are present initially, 
which are weak compared to the super-site tunneling $U_{a,b} \ll |J^{\s}_c|$, 
they cannot cause significant mixing of the symmetric and anti-symmetric super-site orbitals.
Projecting them to the lower band yields a new term 
\begin{equation}
\hat{H} = \sum_j \frac{1}{2} U^* \hat{n}^h_{j,+} ( \hat{n}^h_{j,+} - 1) + \delta \mu \sum_j \hat{n}^h_{j,+}
\end{equation}
with an effective Hubbard interaction $U^\star = (U_a + U_b)/4$ 
and renormalization of the chemical potential by $\delta \mu = - (U_a + U_b)/8$. 
Thus, if in addition to $U_{a,b} \ll |J_c|$ it also holds $|J^{\s}_{a}| \ll U^\star$, 
we can treat the resulting model by assuming that $\hat{h}_{j,+}$ are hard-core bosons. 

\begin{figure}
\includegraphics{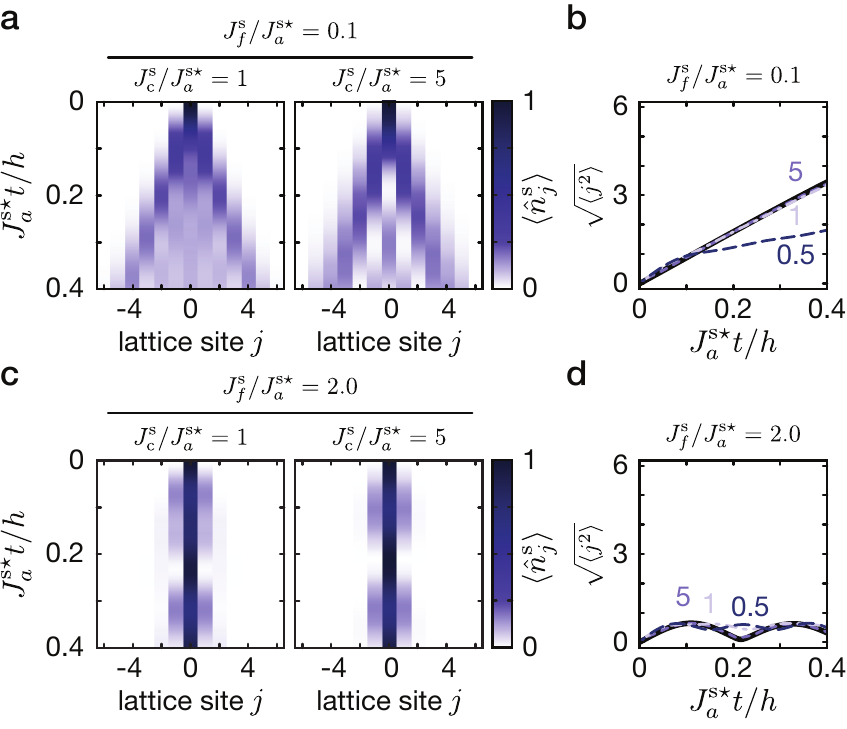}
\caption{\textbf{Numerical time evolution of the 1D \Ztwo{} super-site chain
with one matter particle.}
Dynamics of the matter particle calculated using exact diagonalization 
of a system with 12 \Ztwo{} building blocks connected by 11 super-sites based on Eq.~\eqref{eq:suppl:H1Dsupersite},
with $J^{\mathrm{s}\star}_a=J^{\mathrm{s}}_a/2$. 
The initial state is a matter particle localized on site $j\!=\!0$
and all links are initialized in $\tau^x_{\langle j,j+1\rangle}\!=\!+1$.
\textbf{a} Super-site occupation $n^\mathrm{s}_j$
in a regime with low electric field strength~$J^\mathrm{s}_f/J^{\mathrm{s}\star}_a=0.1$
for different values of the super-site tunnel coupling~$J^\mathrm{s}_c$.
\textbf{b} Root-mean-square distance to the initial position~$\sqrt{\left<j^2\right>}$.
The values of $J^\mathrm{s}_c/J^{\mathrm{s}\star}_a$ are indicated next to the curves.
A free expansion is observed already for $J^\mathrm{s}_c/J^{\mathrm{s}\star}_a \gtrsim 1$,
which agrees well with the ideal evolution expected from the ideal 1D model
captured by Eq.~\eqref{eq:H1D} in the main text (black thick line).
\textbf{c-d} Results in the regime $J^\mathrm{s}_f/J^{\mathrm{s}\star}_a=2$
analogous to \textbf{a-b}.
}
\label{fig:suppl:SupersiteNumerics}
\end{figure}

Because the model in Eq.~\eqref{eq:suppl:H1Dsupersite} 
has local \Ztwo{} gauge symmetries on the super-sites, 
the properties of its eigenstates are qualitatively similar 
to those of the simplified one-band model in Eq.~\eqref{eq:suppl:H1DsupersiteSimplifiedFull},
even if $J^{\s}_c$ is comparable to $J^{\s}_{a}$. 
To demonstrate this, we repeated the simulations from Fig.~\ref{fig:1} in the main text, 
but in a more realistic super-site model. 
We compare cases where $J_a$ in the ideal model from the main text 
is equal to $J^{\s\star}_a$ in the super-site Hamiltonian. 
We find similar behavior for different values of the super-site coupling $J^{\s}_c$ 
(Fig.~\ref{fig:suppl:SupersiteNumerics})
after releasing a \Ztwo{} charge on the central super-site.
Already for small values of the super-site coupling $J^{\s}_c=J^{\s}_a $ in the super-site model 
Eq.~\eqref{eq:suppl:H1Dsupersite}, 
we find a good quantitative agreement with the ideal 1D model introduced in Eq.~(1) in the main text.

\begin{figure}
\includegraphics{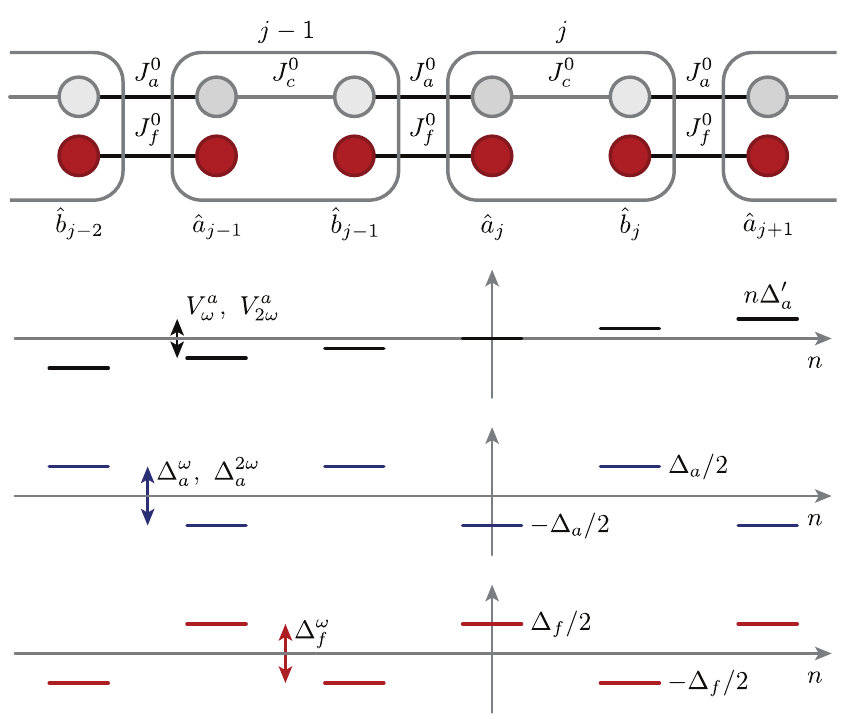}
\caption{\textbf{Floquet implementation of the super-site model.}
The super-sites are illustrated by the gray boxes. 
They consist of $a$ and $b$ sites, 
which are separately illustrated for $a$- and $f$-particles.
The super-site tunnel coupling for the $a$-particles is $J^0_c$
and assumed to be negligible for the $f$-particles.
The $a$-particles are exposed to a linear potential $n\Delta^\prime_a/2$, 
which is modulated by $\omega$ and $2\omega$
with a strength $V_\omega^a$ and $V_{2\omega}^a$.
In addition, they experience a staggered potential 
$\pm\Delta_a$ between neighboring sites, which is also modulated by $\omega$ and $2\omega$
with the strength $\Delta^\omega_a$ and $\Delta^{2\omega}_a$.
The $f$-particles experience an opposite staggered potential with $\Delta_f$.
This potential is modulated with frequency $\omega$ and amplitude $\Delta^\omega_f$.
}
\label{fig:suppl:FloquetImplementationSupersiteModel}
\end{figure}

\textit{Floquet scheme} --- %
The super-site model above can be realized by a Floquet scheme. 
In the following we present one possible microscopic approach,
as illustrated in Fig.~\ref{fig:suppl:FloquetImplementationSupersiteModel}, which is
closely related to the proposal from Ref.~\cite{Barbiero:2018wg:suppl}, 
using the same notations as in Eq.~\eqref{eq:suppl:H1Dsupersite}. 
On the double-well building blocks the processes are identical to the ones 
described in the main text, which we already implemented experimentally. 

We start from the following static Hamiltonian,
\begin{widetext}
\begin{multline}
\hat{H}_0 = - J^0_f \sum_j \hat{\tau}^x_{\langle j+1 , j \rangle} + \frac{\Delta_f}{2} 
\sum_j \hat{\tau}^z_{\langle j+1 , j \rangle} - J^0_a \sum_j \left( \hat{a}_{j+1}^\dagger \hat{b}_j + \hc \right) 
- J^0_c \sum_j \left( \hat{b}_{j}^\dagger \hat{a}_j + \hc \right)  \\
+  \Delta_a' \sum_j \left[ 2 j  \hat{n}^a_j  +  (2 j + 1) \hat{n}^b_j  \right] - \frac{\Delta_a}{2} \sum_j \left[ \hat{n}^a_j - \hat{n}^b_j \right] + \frac{U}{2} \sum_j \left[ (1 - \hat{\tau}^z_{\langle j+1, j \rangle}  ) \hat{n}^b_j + (1 + \hat{\tau}^z_{\langle j+1, j \rangle}  ) \hat{n}^a_{j+1} \right].
\label{eq:suppl:Hmic1}
\end{multline}
\end{widetext}
The first term corresponds to tunneling $J^0_f$ of the $f$-particles between the super-sites, 
which is initially suppressed by the gradient $\Delta_f$. 
The next two terms are bare tunnel couplings of the $a$-particles between 
and within the super-sites. 
They are initially suppressed by the following two terms: 
a linear gradient $\Delta_a'$ and an alternating potential $\pm \Delta_a /2 $ 
on the two inequivalent positions in the super-site. 
The last term corresponds to local inter-species Hubbard interactions 
between $a$ and $f$ particles on all sites.

To obtain the \Ztwo{} gauge invariant super-site model, we supplement $\hat{H}_0$ by the following driving terms,
\begin{widetext}
\begin{multline}
\hat{H}_\omega(t) = \bigl( V_\omega^a \cos (\omega t) + V_{2 \omega}^a \cos (2 \omega t ) \bigr) \sum_j \left[ 2 j  \hat{n}^a_j  +  (2 j + 1) \hat{n}^b_j  \right] +  \frac{\Delta_f^\omega}{2} \cos(\omega t) \sum_j \hat{\tau}^z_{\langle j+1 , j \rangle} \\
- \biggl( \frac{\Delta_a^\omega}{2} \cos(\omega t) + \frac{\Delta_a^{2 \omega}}{2} \cos (2 \omega t) \biggr) \sum_j \left[ \hat{n}^a_j - \hat{n}^b_j \right], 
\label{eq:suppl:Hmic2}
\end{multline}
\end{widetext}
which will be used to restore tunnel couplings of the $a$- and $f$-particles. 
The  first term corresponds to a modulated linear gradient seen by the $a$-particles. 
The other two terms are modulated alternating potentials 
seen by the $f$- and $a$-particles, respectively.
The modulations at two frequencies $\omega$ and $2 \omega$ 
are used to control both the amplitude and phase of the resulting effective Floquet Hamiltonian, 
see Refs.~\cite{Barbiero:2018wg:suppl,Gorg:2018de:suppl}.

Following the scheme proposed in Ref.~\cite{Barbiero:2018wg:suppl}, 
we require the following resonance conditions:
\begin{flalign}
\hbar \omega = \Delta_f &= U \label{eq:resstart}\\
\Delta_a = \Delta_a' &= U/2 \\
\Delta_a^{2\omega} = V_{2 \omega}^a &\equiv V_{2 \omega}^x / 2\\
V_\omega^a - \Delta_a^\omega = \Delta_f^\omega &\equiv V_\omega^y \\
V_\omega^a + \Delta_a^\omega & \equiv V_\omega^x.
\end{flalign}
The last three lines include the three amplitudes $V_{2 \omega}^x$, $V_\omega^x$ 
and $V_\omega^y$ which are fixed by
\begin{flalign}
x^{(1)} = V_\omega^x / (\hbar \omega) &\simeq 1.71\\
x^{(2)} = V_{2 \omega}^x / (\hbar \omega) &\simeq 1.05\\
y^{(1)} = V_\omega^y / (\hbar \omega) &\simeq 1.84 
\label{eq:resfinal}
\end{flalign}
These are solutions to $\mathcal{J}_0(y^{(1)}) = \mathcal{J}_2(y^{(1)})$ 
and $\lambda_0 = \lambda_1 = \lambda_2 \equiv \lambda_{012}$ 
where $\lambda_n = \sum_{\ell 
= - \infty}^\infty \mathcal{J}_{n-2 \ell}(x^{(1)}) \mathcal{J}_\ell(x^{(2)}/2)$.

In the large frequency limit, and under the above conditions, 
the effective Floquet Hamiltonian becomes \cite{Goldman:2015kca:suppl}
\begin{multline}
\hat{H}_{\rm eff} = - J^0_f  \sum_j \hat{\tau}^x_{\langle j+1 , j \rangle} 
\hat{\Lambda}^y_{\langle j+1 , j \rangle} - J^0_c \sum_j \lambda^x 
\left( \hat{a}^\dagger_{j} \hat{b}_j + \hc\right) \\
- J^0_a \sum_j \lambda^y \left(\hat{a}_{j+1}^\dagger \hat{\tau}^z_{\langle j+1,j \rangle} 
\hat{b}_{j} + \hc \right).
\label{eqHeff}
\end{multline}
The renormalization of tunneling amplitudes is given by
\begin{multline}
\hat{\Lambda}^y_{\langle j+1 , j \rangle} = \frac{1}{2} 
\left( 1 - (-1)^{\hat{n}^a_{j+1}+\hat{n}^b_{j}} \right) \mathcal{J}_0(y^{(1)} ) \\
+\frac{1}{2} \left( 1 + (-1)^{\hat{n}^a_{j+1}+\hat{n}^b_{j}} \right) \mathcal{J}_1(y^{(1)} ),
\end{multline}
$\lambda^x=\lambda_{012} \simeq 0.37$ and $\lambda^y = \mathcal{J}_1(y^{(1)}) \simeq 0.58$.

One easily confirms that Hamiltonian \eqref{eqHeff} 
is \Ztwo{} gauge invariant if the super-site gauge operator is used,
\begin{equation}
\hat{G}_{j,ab} = \hat{Q}_{j,ab} \hat{\tau}^x_{\langle j-1 , j \rangle} 
\hat{\tau}^x_{\langle j , j+1 \rangle}.
\end{equation}
The effective Hamiltonian \eqref{eqHeff} 
is similar to the super-site model in Eq.~\eqref{eq:suppl:H1Dsupersite}, 
except for the fact that the renormalization of the $f$-particle tunneling, 
$\hat{\Lambda}^y_{\langle j+1 , j \rangle}$, is operator-valued 
and depends on the \Ztwo{} charges on the adjacent sites. 
This is not expected to change the physics of the model. 
If required, two-frequency driving can also be used for the $f$-particles 
to obtain a situation where $\hat{\Lambda}^y \to \Lambda^y$ becomes a $\mathbb{C}$ number. 

\begin{figure*}
\includegraphics{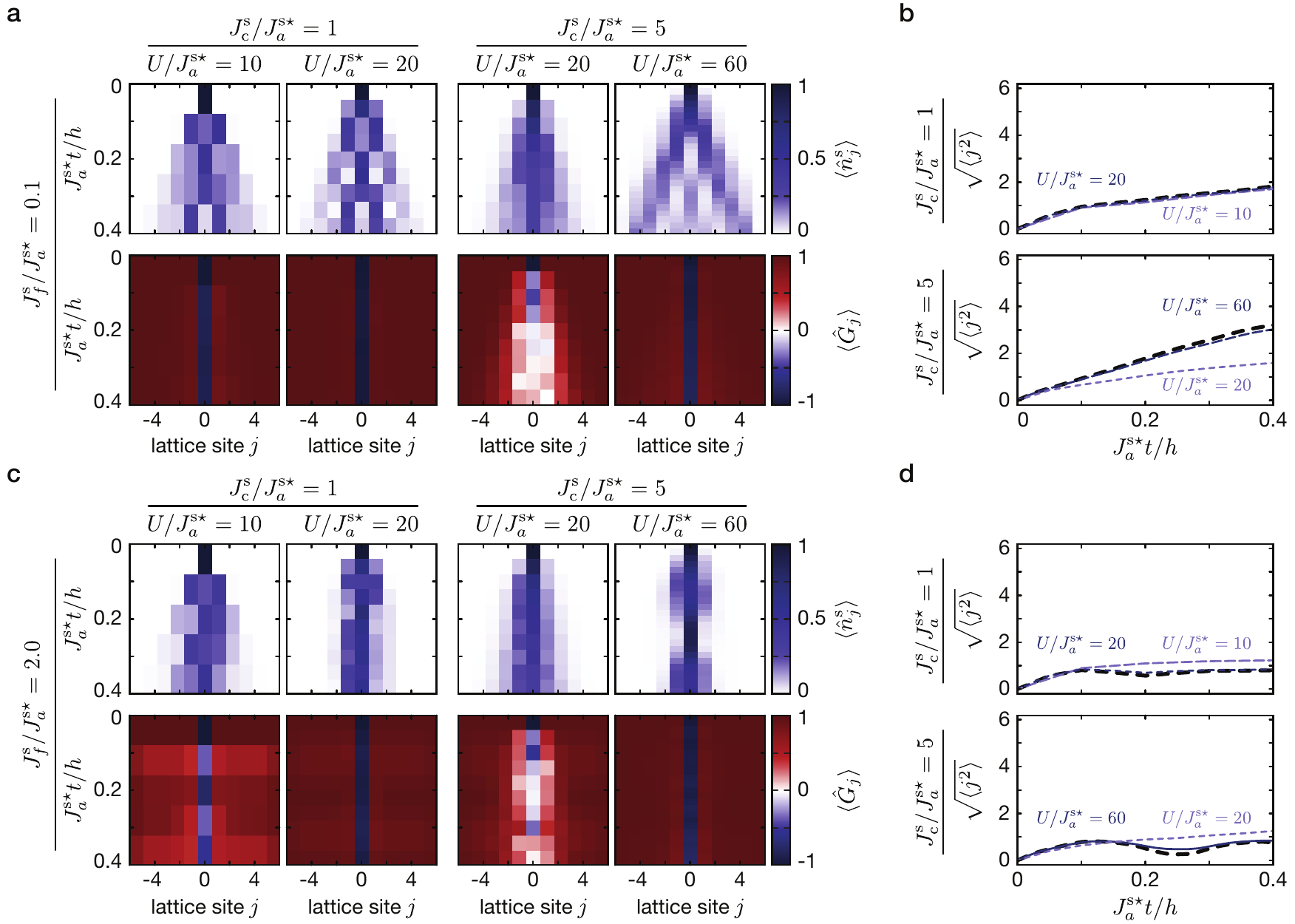}
\caption{\textbf{Numerical time evolution of the full time-dependent model with one particle.}
Dynamics of the matter particle calculated by numerically solving the time-dependent model
according to Eqs.~\eqref{eq:suppl:Hmic1} and~\eqref{eq:suppl:Hmic2}, fulfilling the resonance conditions
Eqs.~\eqref{eq:resstart} -- \eqref{eq:resfinal} and setting $J_a^0=2J_a^{\mathrm{s}\star}/\mathcal{J}_1(y^{(1)})$, $J_c^0=J_c/\lambda^x$ and $J_f^0=J_f/\mathcal{J}_1(y^{(1)})$.
The system consists of 10 \Ztwo{} building blocks and 9 super-sites.
The edges of the 1D chain have therefore only a single site. 
The initial state is a matter particle localized on site $j\!=\!0$
and all links are initialized in $\tau^x_{\langle j,j+1\rangle}\!=\!+1$.
\textbf{a} Matter particle dynamics 
and expectation value of the local symmetry operator~$\langle\hat{G}_j\rangle$ 
in a regime with low electric field strength~$J^\mathrm{s}_f/J^{\mathrm{s}\star}_a=0.1$
for different inter-species interaction energies~$U$.
\textbf{b} 
Root-mean-square distance to the initial position~$\sqrt{\left<j^2\right>}$
of the matter particle dynamics from \textbf{a} 
in comparison to the high-frequency limit.
The results in the high-frequency limit are shown by thick dashed black lines.
\textbf{c-d} Results in the regime $J^\mathrm{s}_f/J^{\mathrm{s}\star}_a=2$
analogous to~\textbf{a-b}.
}
\label{fig:suppl:FloquetSimulationSupersiteModel-1P}
\end{figure*}

\begin{figure*}
\includegraphics{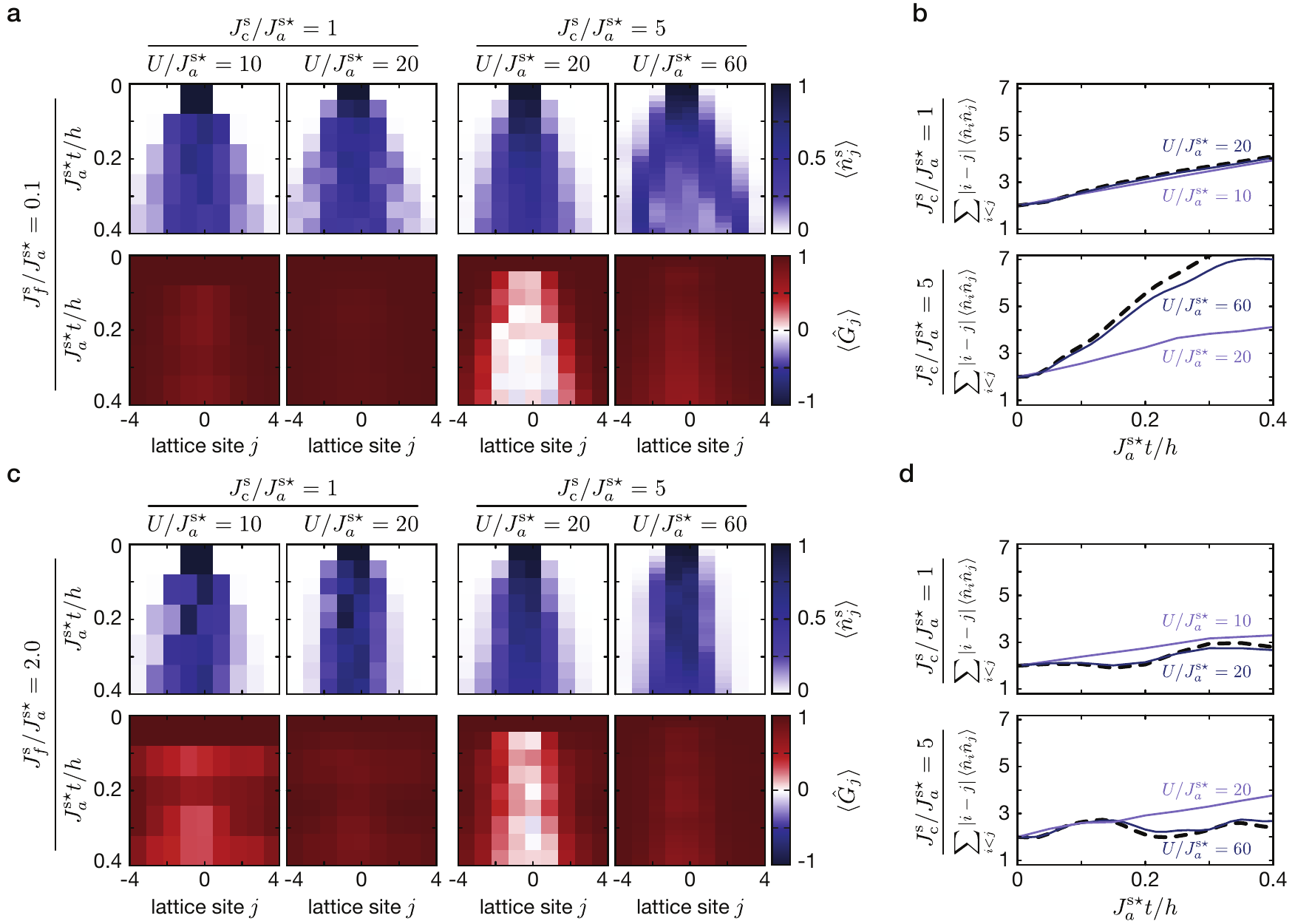}
\caption{\textbf{Numerical time evolution of the full time-dependent model with two particles.}
Dynamics of two matter particles calculated by numerically solving the time-dependent model
according to Eqs.~\eqref{eq:suppl:Hmic1} and~\eqref{eq:suppl:Hmic2}, fulfilling the resonance conditions
Eqs.~\eqref{eq:resstart} -- \eqref{eq:resfinal} and setting $J_a^0=2J_a^{\mathrm{s}\star}/\mathcal{J}_1(y^{(1)})$, $J_c^0=J_c/\lambda^x$ and $J_f^0=J_f/\mathcal{J}_1(y^{(1)})$.
The system consists of 8 \Ztwo{} building blocks and 9 super-sites.
The system's edge is therefore a super-site.
The initial state is a matter particle localized on site $j\!=\!0$
and a second matter particle at site $j\!=\!-1$.
The link between the two sites is $\tau^x_{\langle -1,0\rangle}\!=\!-1$.
All other links are initialized in $\tau^x_{\langle j,j+1\rangle}\!=\!+1$.
\textbf{a} Matter particle dynamics 
and expectation value of the local symmetry operator~$\langle\hat{G}_j\rangle$ 
in a regime with low electric field strength~$J^\mathrm{s}_f/J^{\mathrm{s}\star}_a=0.1$
for different inter-species interaction energies~$U$.
\textbf{b} 
Time evolution of the distance between the two particles 
measured by $\sum_{i<j}|i-j|\langle\hat{n}_i\hat{n}_j\rangle$
for the data shown in \textbf{a},
in comparison to the high frequency-limit.
The results in the high-frequency limit are shown by thick black lines.
\textbf{c-d} Results in the regime $J^\mathrm{s}_f/J^{\mathrm{s}\star}_a=2$
analogous to~\textbf{a-b}.
}
\label{fig:suppl:FloquetSimulationSupersiteModel-2P}
\end{figure*}

To illustrate the accuracy of our Floquet scheme, 
in Fig.~\ref{fig:suppl:FloquetSimulationSupersiteModel-1P} we repeat our calculations 
from Fig.~\ref{fig:suppl:SupersiteNumerics} starting from the microscopic model.
The parameters in Eqs.~\eqref{eq:suppl:Hmic1}, 
\eqref{eq:suppl:Hmic2} are chosen such 
that the effective couplings in Eq.~\eqref{eqHeff} correspond to the parameters 
we used in the super-site model Eq.~\eqref{eq:suppl:H1Dsupersite}. 
For $U = 10 J^0_a$ in the microscopic Hamiltonian, 
we find that the \Ztwo{} gauge invariance remains valid for long times 
and the observed dynamics is similar to our expectation from the super-site model. 
In Fig.~\ref{fig:suppl:FloquetSimulationSupersiteModel-2P} we also performed simulations 
starting from an initial state with two fermionic \Ztwo{} charges $a$ located next to each other, 
connected by a \Ztwo{} electric field line. 
In this case, too, the gauge invariance of the model remains intact for long times. 
This establishes the proposed Floquet scheme as a realistic method 
to implement and realize dynamics in models with local \Ztwo{} gauge constraints.

---

\end{document}